# Influence of interfacial slip on the suspension rheology of a dilute emulsion of surfactant-laden deformable drops in linear flows

Sayan Das, Anirban Bhattacharjee and Suman Chakraborty†

Department of Mechanical Engineering, Indian Institute of Technology Kharagpur,
Kharagpur – 721302, India

The present study deals with the effect of interfacial slip on the deformation and emulsion rheology of a dilute suspension of droplets in a linear flow. The droplets are laden with surfactants that are bulk-insoluble and get transported only along the interface. An asymptotic approach is adopted for the present analysis in order to tackle the nonlinearity present due to deformation of droplets. The analysis is carried out under two different limiting scenarios namely: surface diffusion-dominated-surfactant transport and surface convection-dominated surfactant transport. For either of the limiting cases we look into the droplet dynamics for two commonly encountered bulk flows - uniaxial extensional and simple shear flow. Under the assumption of negligible fluid inertia in either phase, it is shown that slip at the droplet interface significantly affects the surfactant-induced Marangoni stress and hence droplet deformation and emulsion rheology. Presence of interfacial slip not only brings about a decrease in the droplet deformation but also reduces the effective viscosity of the emulsion. The fall in both droplet deformation and effective viscosity is found to be more severe for the limiting case of surface convection-dominated surfactant transport. For the case of an imposed simple shear flow, the normal stress differences generated due to droplet deformation are affected as well due to the presence of interfacial slip.

† E-mail address for correspondence: suman@mech.iitkgp.ernet.in

## I. INTRODUCTION

The study of droplet dynamics, among other fields, has been a keen area of interest due to its wide application in not only conventional industrial processes but also in various microfluidic applications involving the generation, transport and deformation of droplets through different microfluidic devices that is required for drug delivery processes as well as cell encapsulation, analytic detection, mixing of reagents etc. [1–4] Intermixing of two different phases is quite common in various emulsifications and blending processes, such as fiber spinning, blow molding, biaxial stretching or thermo-forming, where one of the phases is suspended in the form of droplets. On application of external forces to the emulsions, the droplets deform, the study of which bears significant importance in the field of rheology of various emulsions [5–8]. It has been proven that properties of emulsions are dependent on the concentration as well as on the morphology of the dispersed phase [8,9]. For instance, shape deformations of droplets in shear flow can generate first and second normal stress differences that give rise to rheological behavior of emulsions similar to polymeric fluids [9,10]. Hence an understanding of the emulsion rheology and the deformation of suspended droplets serves as an important prerequisite to the design and manipulation of various emulsification processes in industries.

Since the classical study done by Taylor [11], several analyses have been performed on the droplet hydrodynamics in linear flows. Out of these, a significant number took into account the effect of surfactants or contaminants on droplet deformation as well as on emulsion rheology [12–14]. For example, Stone and Leal [13], used a small deformation asymptotic theory to analyze the role played by bulk-insoluble surfactants on droplet deformation in the limiting case of diffusion-dominated surfactant transport. They also verified their theoretical results numerically. Further, Milliken et al. [15] numerically investigated the deformation and breakup of a surfactant-laden droplet in a uniaxial extensional flow field. Several experimental studies have provided evidence regarding the existence of a relationship between droplet deformation and local surfactant concentration [13,15–20]. A large number of numerical studies also have been carried out to analyze the droplet transient deformations as well as break-up due to the presence of surfactants along the droplet surface [13,16,21–23].

Although most of the studies undertaken till date describe the fluid-fluid interface in a multiphase system by the presence of surface tension and curvature, recent experiments on different polymers report the presence of an apparent slip along the interfaces [24–26]. The origin of this slip is due to a decrease in the viscosity at the interface [27]. From a microscopic point of view, the fluid-fluid interface acts as a diffuse region where a partial mixing of either of the phases takes place due to which the degree of entanglement severely decreases as compared to that in the bulk [27,28]. Different molecular dynamics simulations and experiments have provided evidence of augmentation in slip along the interface between two non-Newtonian fluids with the use of hydrophobic beads [29]. Hence the fluid-fluid interface may possess slip in velocity due to a fall in interfacial viscosity for polymers or due to the presence of surfactants. A number of recent studies deals with the effect of interfacial slip on the droplet dynamics [9,30–32]. Ramachandran and Leal[9] studied the effect of interfacial slip on the rheology and dynamics of a dilute emulsion of droplets suspended in linear flows. They showed that increase in the slip at the interface reduces both the droplet deformation as well as the effective viscosity of the emulsion.

Recently, Vlahovska et al. and Mandal et al.[12,14] investigated the effect of surfactants on the deformation characteristics as well as the emulsion rheology of a dilute suspension of droplets in linear flows. However, there is no study available in the literature that puts forward the effect of interfacial slip on the dynamics of a surfactant-laden droplet. Surfactants at the droplet surface bring in alterations in the interfacial rheology that promotes the presence of interfacial slip, which is quite realistic in any multiphase system and is found to possess a significant effect on droplet dynamics [9,31]. This is the prime objective of the present study. It is seen that interfacial slip induces a greater stability to the emulsion by reducing the droplet deformation brought about by the imposed linear flow. Also, it reduces the effective viscosity of the suspension. It is observed that there is a good match between our theoretical prediction and previously performed experiments when the effect of interfacial slip is taken into consideration in our analysis. For the limiting case of convection dominated surfactant transport, the influence of slip is found to be highly effective.

## II. THEORETICAL MODEL

### A. System Description

The present system consists of a neutrally buoyant Newtonian droplet (of radius $a$ and viscosity $\mu_i$) suspended in another Newtonian fluid (of viscosity $\mu_e$) in the presence of an imposed linear flow, namely, a simple shear flow or a uniaxial extensional flow. The density of either of the phases is taken to be $\rho$. Surfactants are assumed to be insoluble in either of the phases, that is they are transported along the surface of the droplet. The subscript '$i$' is used to indicate the phase inside the droplet whereas the subscript '$e$' refers to the carrier phase. In the absence of any imposed fluid flow, the surfactant is uniformly distributed along the droplet surface with its concentration denoted by $\bar{\Gamma}_{eq}$ and the corresponding surface tension is denoted by $\bar{\sigma}_{eq}$. For the case of a surfactant-free system, we denote the equilibrium surface tension by $\bar{\sigma}_c$. Presence of imposed flow, however, disturbs this equilibrium due to initiation of fluid flow along the droplet surface. The resulting surfactant concentration and its corresponding surface tension are denoted by $\bar{\Gamma}$ and $\bar{\sigma}$ respectively. This variation of surface tension along the droplet surface gives birth to the Marangoni stress that has a significant role in altering the droplet dynamics[12,13,33]. In addition, the presence of interfacial slip renders the tangential component of velocity at the interface of the droplet discontinuous. The imposed linear flow, in general, is denoted by $\bar{\mathbf{u}}_\infty$ and can be expressed as

$$\bar{\mathbf{u}}_\infty = \left(\bar{\mathbf{D}}_\infty + \bar{\mathbf{\Omega}}_\infty\right)\cdot \bar{\mathbf{x}}, \tag{1}$$

where $\bar{\mathbf{D}}_\infty$ and $\bar{\mathbf{\Omega}}_\infty$ are the velocity gradient tensor and the vorticity tensors respectively and $\bar{\mathbf{x}}$ is the position vector. For the case of a uniaxial-extensional flow, both the above tensors can be expressed as [9]

$$\bar{\mathbf{D}}_\infty = \frac{\dot{\gamma}}{2}\begin{bmatrix} -1 & 0 & 0 \\ 0 & -1 & 0 \\ 0 & 0 & 2 \end{bmatrix}, \quad \bar{\mathbf{\Omega}} = \mathbf{0}. \tag{2}$$

whereas for imposed simple shear flow we have [9]

$$\bar{\mathbf{D}}_\infty = \frac{\dot{\gamma}}{2}\begin{bmatrix} 0 & 1 & 0 \\ 1 & 0 & 0 \\ 0 & 0 & 0 \end{bmatrix}, \quad \bar{\mathbf{\Omega}} = \frac{\dot{\gamma}}{2}\begin{bmatrix} 0 & 1 & 0 \\ -1 & 0 & 0 \\ 0 & 0 & 0 \end{bmatrix}, \tag{3}$$

where $\dot{\gamma}$ is the rate of shear or extension. It should be noted that all the quantities bearing a 'bar' are dimensional quantities, whereas those without any 'bar' signify dimensionless quantities.

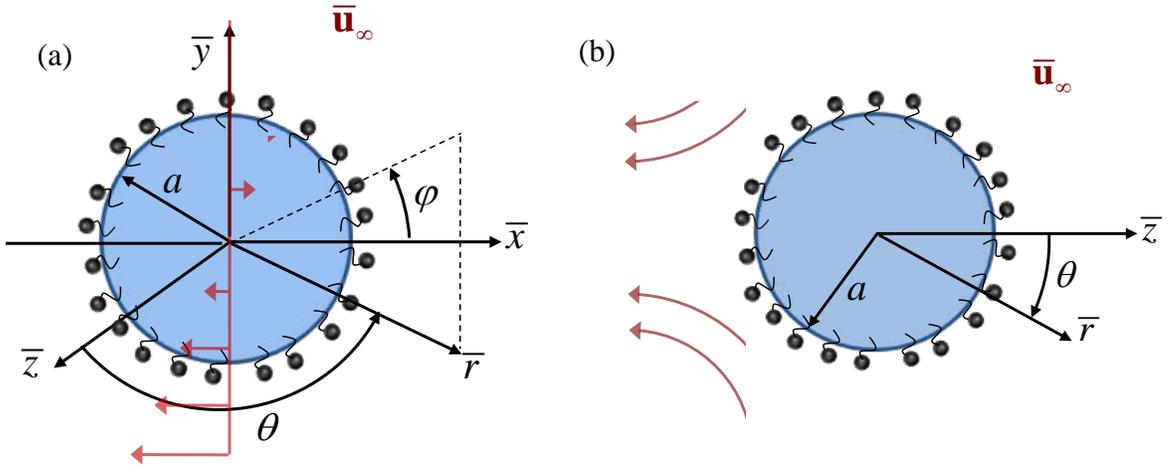

Fig. 1: Schematic of a surfactant-laden droplet of radius $a$ suspended in a linear flow. In fig. (a) the background flow is shown as a simple shear flow whereas in fig. (b) the imposed flow is displayed as a uniaxial extensional flow. Both spherical $(\bar{r},\theta,\varphi)$ and cartesian coordinates $(\bar{x},\bar{y},\bar{z})$ are shown, which are fixed to the centroid of the droplet.

A schematic of the of the physical system under consideration is shown in fig. 1. Figure 1(a) shows the case when the imposed flow is a simple shear flow, whereas in fig. 1(b) the droplet is suspended in an axisymmetric extensional flow. In either of the cases, the spherical coordinate $(\bar{r}, \theta, \varphi)$ is attached to the centroid of the droplet. The primary aim of our study is to investigate the role played by interfacial slip on the deformation of a surfactant-laden droplet suspended in a linear flow under two limiting scenarios: (i) when surfactant transport is dominated by surface diffusion and (ii) when surface convection is the main mode of surfactant transport. In addition, we also analyze the effect of interfacial slip on the emulsion rheology for either type of linear flows.

## B. Governing equations and boundary conditions

Prior to stating the governing differential equations and relevant boundary conditions, we put forward some important assumptions required to simplify them so that they are analytically solvable. These assumptions include the following: (*i*) negligible fluid inertia so that the fluid flow hydrodynamics falls in the creeping flow or low Reynolds number regime $\left(Re = \rho \dot{\gamma} a^2 / \mu_e \ll 1\right)$, (*ii*) small deviations from the spherical shape of the droplet, which is possible only when the interfacial tension along the droplet surface dominates the viscous forces acting on the droplet, thus indicating that the capillary number is low $\left(Ca^* = \mu_e \dot{\gamma} a / \bar{\sigma}_c \ll 1\right)$. The capillary number signifies the relative strength of the viscous force acting on the droplet as compared to the surface tension force along its interface. Some other assumptions are: (*iii*) the surfactants are considered to be insoluble in either of the phases and are transported along the droplet surface solely, (*iv*) for a dilute concentration of surfactants along the droplet surface, the surface tension is assumed to bear a linear relationship with the surfactant distribution, and finally (*v*) the droplet dynamics is assumed to be unaffected due to the presence of any bounding walls.

The flow field and the surfactant transport along the droplet surface are governed by the Navier-Stokes equation and convection-diffusion equation respectively. Under the premise of the above assumptions, the governing equation of flow field reduces to the Stokes equation which is subjected to kinematic and stress balance conditions at the interface. In addition, the flow field inside the droplet is bounded at the centroid and outside the droplet it satisfies the far-field condition given by the imposed flow. Due to the consideration of convection of surfactants along the surface of the droplet, the surfactant transport equation, which governs the surfactant concentration, is coupled with the flow field. We first specify the following scheme used to non-dimensionalize the governing equations and boundary conditions: $r = \bar{r}/a$, $\mathbf{u} = \bar{\mathbf{u}}/\dot{\gamma} a$, $\Gamma = \bar{\Gamma}/\bar{\Gamma}_{eq}$, $\sigma = \bar{\sigma}/\bar{\sigma}_c$, $p = \bar{p}/(\mu_e \dot{\gamma})$, $\boldsymbol{\tau} = \bar{\boldsymbol{\tau}}/(\mu_e \dot{\gamma})$. The different non-dimensional parameters and numbers that are obtained while deriving the non-dimensional governing equations and boundary conditions are (*i*) viscosity contrast or the viscosity ratio,

$\left( \lambda = \mu_i / \mu_e \right)$ defined as the ratio of the viscosity of the droplet phase to the carrier phase, (*ii*) the elasticity number $\left( \beta = \bar{\Gamma}_{ref} R \bar{T}_o / \bar{\sigma}_c = -d\left( \bar{\sigma} / \bar{\sigma}_c \right) / d\bar{\Gamma} \right)$, which is an indicator of the sensitivity of the surface tension towards the surfactant concentration along the droplet interface, (*iii*) the surface Péclet number $\left( Pe_s = \dot{\gamma} a^2 / D_s \right)$ that signifies the relative importance of surfactant transport by convection in comparison to that by diffusion along the droplet surface, ($D_s$ is the surface diffusivity of the surfactants) (*iv*) the modified capillary number, $Ca = Ca^* / (1 - \beta)$. The choice of such a modified capillary number is solely for the purpose of ease in calculation[14]. The modified capillary number is based on the equilibrium surface tension for a droplet uniformly coated with surfactants $\left[ \bar{\sigma}_{eq} = \bar{\sigma}_c (1 - \beta) \right]$.

With the aid of the above non-dimensional scheme, we now obtain the dimensionless set of governing differential equations and boundary conditions. The continuity and the Stokes equation that govern the flow, both inside as well as outside the droplet, can be written as

$$\nabla \cdot \mathbf{u}_{i,e} = 0, \quad -\nabla p_{i,e} + \lambda \nabla^2 \mathbf{u}_{i,e} = \mathbf{0}, \tag{4}$$

where $\mathbf{u}_{i,e}$ is the velocity field and $p_{i,e}$ refers to the pressure field. The above set of equations are subjected to the following boundary conditions

$$
\begin{aligned}
&(i) \quad \text{At } r \to \infty, \left( \mathbf{u}_e, p_e \right) \to \left( \mathbf{u}_\infty, p_\infty \right), \\
&(ii) \quad \text{At } r = 0, \left( \mathbf{u}_i, p_i \right) \text{ are bounded}, \\
&(iii) \quad \text{At } r = r_s, \quad \mathbf{u}_i \cdot \mathbf{n} = \mathbf{u}_e \cdot \mathbf{n} = \mathbf{0}, \\
&\qquad \qquad \qquad \mathbf{u}_e = \mathbf{u}_i + \delta (\mathbf{I} - \mathbf{nn}) \cdot \boldsymbol{\tau}_e \cdot \mathbf{n}, \\
&\qquad \qquad \qquad \boldsymbol{\tau}_e \cdot \mathbf{n} - \boldsymbol{\tau}_i \cdot \mathbf{n} = \left( \frac{\beta}{1 - \beta} \right) \frac{1}{Ca} \nabla_s \Gamma + \frac{\sigma}{Ca} (\nabla \cdot \mathbf{n}),
\end{aligned}
\tag{5}
$$

where $\left( \mathbf{u}_\infty, p_\infty \right)$ are the non-dimensional velocity and pressure at the far-field. The dimensionless velocity at the far-field can be obtained from equation (1). Amongst the above set of equations, the boundary condition (*i*) ensures the fact that the flow field outside the droplet satisfies the far-field condition which is set by the imposed flow. The boundary condition (*ii*) puts forward the boundedness of both the velocity and the pressure field at the centroid of the droplet. The rest of the boundary conditions (*iii*) are applied at the droplet surface $(r = r_s)$, where $r_s$ is the radial position of the droplet interface. The first among them is the impermeability condition which signifies that either of the phases involved in the analysis is immiscible. The second boundary condition brings into picture the jump in the velocity field across the interface due to the presence of slip[31,32]. Here $\delta$ is the slip coefficient that indicates the amount of the slip in velocity. The slip coefficient can be represented, from a microscopic point of view as $\delta = h_i/\eta_i$, where $h_i$ is the thickness of the interface and $\eta_i$ represents its viscosity[9,30]. The final boundary condition in equation (5) is the stress balance along the interface.

The deformed droplet surface can be represented as $r_s = 1 + g(\theta, \varphi)$ where $g(\theta, \varphi)$ represents the correction to the spherical droplet shape. $\nabla_s = (\mathbf{I} - \mathbf{nn}) \cdot \nabla$ indicates the surface gradient tensor, where $\mathbf{n}$ is the unit normal drawn to the deformed surface of the droplet and can be expressed as $\mathbf{n} = \nabla F / |\nabla F|$. Here $F = r = r_s$ represents the equation of the surface of the droplet. In the stress balance condition, $\boldsymbol{\tau}_{i,e}$ is the hydrodynamic stress tensor inside and outside the droplet, given by $\boldsymbol{\tau}_i = -p_i \mathbf{I} + \lambda \left[ \nabla \mathbf{u}_i + (\nabla \mathbf{u}_i)^T \right]$ and $\boldsymbol{\tau}_e = -p_e \mathbf{I} + \left[ \nabla \mathbf{u}_e + (\nabla \mathbf{u}_e)^T \right]$ [34]. The stress balance condition in equation (5) is obtained as a result of substitution of following equation of state relating the surface tension with the surfactant concentration along the droplet interface [13,14,35]

$$\sigma = 1 - \beta \Gamma \tag{6}$$

The above surface tension is non-dimensionalized with respect to the equilibrium surface tension of a droplet uniformly coated with surfactant, that is, $\sigma = \bar{\sigma}/\bar{\sigma}_{eq} = \bar{\sigma}/(1-\beta)\bar{\sigma}_c$ [13,14]. It is thus obvious that $0 < \beta < 1$. The governing equation for the transport of surfactants along the interface, keeping in mind the insolubility of surfactants in either of the phases, is given by

$$Pe_s \nabla_s \cdot (\mathbf{u}_s \Gamma) = \nabla_s^2 \Gamma. \tag{7}$$

The above equation is a simplified convection-diffusion equation for the special case of dilute concentration of surfactants on the droplet surface. While solving for the flow field and the surfactant concentration, it is important to check for the mass conservation of the surfactants along the droplet surface. This constraint for conservation of mass can be mathematically represented as follows

$$\int_{\varphi=0}^{2\pi} \int_{\theta=0}^{\pi} \Gamma(\theta, \varphi) r_s^2 \sin\theta \, d\theta \, d\varphi = 4\pi. \tag{8}$$

On observing equations (4), (5) and (7), it can be inferred that the equations governing the flow field and the relevant boundary conditions are coupled with the governing equation for surfactant transport. In addition to this the shape of the deformed droplet is also unknown, which has to be obtained as a portion of the solution of the flow field. Hence an exact analytical solution for arbitrary values of $Ca$ and $Pe_s$ is impossible. To tackle this issue, we adopt an asymptotic methodology for small droplet deformation $(Ca \ll 1)$ under the limiting case of surfactant transport primarily governed by surface diffusion $(Pe_s \ll 1)$ and by surface convection $(Pe_s \to \infty)$ [14,36].

## III. ASYMPTOTIC APPROACH

In the limiting case of low $Pe_s$, surface diffusion of surfactants is the primary mode of surfactant transport along the droplet interface. Since only small deformation of the droplet is taken into consideration, it is worthwhile to assume that $Pe_s \sim Ca$ or

$$Pe_s = kCa, \tag{9}$$

where $k = a\bar{\sigma}_c(1-\beta)/\mu_e D_s$ is a finite constant of magnitude of $O(1)$. This constant is known as the property parameter as it is dependent on different material properties. Thus it can be rightfully justified that the droplet deformation is a function of the parameters $k, \delta, \beta$ and $\lambda$. The surfactant transport equation in the limiting case of $Pe_s \ll 1$, can be expressed as[14]

$$kCa \nabla_s \cdot (\mathbf{u}_s \Gamma) = \nabla_s^2 \Gamma. \tag{10}$$

Since small deformation of the droplet is assumed, $Ca$ is thus an appropriate candidate for the perturbation parameter. Hence any flow field variable (say $\psi$) can be expanded in increasing powers of $Ca$ as follows[14,36]

$$\psi = \psi^{(0)} + \psi^{(Ca)} Ca + O(Ca^2), \tag{11}$$

where $\psi^{(0)}$ is the leading order term corresponding to no deformation, whereas the term $\psi^{(Ca)}$ signifies the $O(Ca)$ correction term due to the presence of droplet deformation. Any other terms on the right indicate even higher order corrections to the flow variable due to shape deformation. Keeping in mind the total mass conservation constraint on the droplet surface in equation (8), the surfactant concentration is expanded in the following manner[14]

$$\Gamma = 1 + \Gamma^{(0)} Ca + \Gamma^{(Ca)} Ca^2 + O(Ca^3). \tag{12}$$

All the flow variables involved are expressed in terms of spherical harmonics at each level of perturbation in our analysis. For instance, the surfactant concentration can be expressed as[14,36]

$$\Gamma = \sum_{n=0}^{\infty} \sum_{m=0}^{n} \left[ \Gamma_{n,m} \cos(m\varphi) + \hat{\Gamma}_{n,m} \sin(m\varphi) \right] P_{n,m}(\cos\theta), \tag{13}$$

where $P_{n,m}(\cos\theta)$ represents an associated Legendre polynomial of order $m$ and degree $n$. The constant coefficients, $\Gamma_{n,m}$ and $\hat{\Gamma}_{n,m}$ are unknown and are found out as a part of the solution. In the present study, we take help of the Lamb's general solution for Stokes equation given in (4) for either of the phases involved. We follow the same methodology as was used by Mandal et al.[14] to obtain the solution for the flow field and the surfactant concentration at each level of perturbation. The detailed expression of the Lamb's general solution for velocity and pressure field can be found in the same study. In order to first obtain the leading order solution, all the flow field boundary conditions at the undeformed droplet surface (kinematic and tangential stress balance condition) are solved simultaneously with the leading order surfactant transport equation. With the leading order solution at our disposal, we make use of the leading order normal stress boundary condition to obtain the $O(Ca)$ correction to the shape. The tangential stress balance can be obtained from equation (5) and expressed in the following manner [14,34,36]

$$\text{at } r = r_s, \quad (\boldsymbol{\tau}_e \cdot \mathbf{n} - \boldsymbol{\tau}_i \cdot \mathbf{n}) \cdot (\mathbf{I} - \mathbf{nn}) = \frac{\beta}{(1-\beta)Ca}(\nabla_s \Gamma) \cdot (\mathbf{I} - \mathbf{nn}), \tag{14}$$

where $\mathbf{I}$ is the identity tensor and $r_s = 1 + Ca g^{(Ca)} + Ca^2 g^{(Ca^2)} + O(Ca^3)$ is the radial position of the deformed surface of the droplet., where $O(Ca)$ correction to the droplet shape is given by

$$g^{(Ca)} = \sum_{n=0}^{\infty}\sum_{m=0}^{n}\left[L_{n,m}^{(Ca)}\cos(m\varphi) + \hat{L}_{n,m}^{(Ca)}\sin(m\varphi)\right]P_{n,m}(\cos\theta). \tag{15}$$

Here $L_{n,m}^{(Ca)}$ and $\hat{L}_{n,m}^{(Ca)}$ are unknown constant coefficients which can be found out from the normal stress balance. Thus with the leading order solution at our disposal, we make use of the leading order normal stress balance to obtain the $O(Ca)$ correction in droplet shape. The normal stress balance, in general, can be expressed as [33,34,36]

$$\text{at } r = r_s, \quad (\boldsymbol{\tau}_e \cdot \mathbf{n} - \boldsymbol{\tau}_i \cdot \mathbf{n}) \cdot \mathbf{n} = \frac{\sigma}{Ca}(\nabla \cdot \mathbf{n}) \tag{16}$$

In a similar manner $g^{(Ca^2)}$ can be expressed in terms of spherical harmonics. With the aid of leading order solution for flow field and surfactant concentration as well as the $O(Ca)$ shape correction, we solve for the $O(Ca)$ flow field followed by the $O(Ca^2)$ shape deformation. The $O(Ca)$ solution is obtained by again simultaneously solving the flow field boundary conditions and the surfactant transport equation derived along the deformed droplet surface. We finally use the $O(Ca)$ normal stress balance to obtain the $O(Ca^2)$ shape deformation of the droplet

$\left[ g^{(Ca^2)} \right]$ which can be similarly be expressed in terms of spherical harmonics. For details of the asymptotic approach, one can refer to the work of Mandal et al.[14].

Under the limiting case of high surface Péclet number, the sole difference in the approach lies in the surfactant transport equation, which is given by

$$\nabla_s \cdot (\mathbf{u}_s \Gamma) = 0. \tag{17}$$

Instead of opting for the above approach to obtain the solution, one apply the limit $k \to \infty$ to all the solutions obtained in the previous limiting case. This is the strategy that we have followed to obtain the solution for the high $Pe_s$ limit. We now highlight the important results obtained from the analysis for the two different type of linear flows considered in this study, namely, uniaxial-extensional flow and simple shear flow.

## A. Uniaxial Extensional flow

A schematic for the case when the imposed flow is a uniaxial-extensional flow, is provided in figure 1(b). The flow field and the surfactant concentration are solved at different orders of perturbation. As discussed before, the leading order surfactant concentration is obtained by simultaneously solving the flow field boundary conditions and the surfactant transport equation for the same order and can be expressed as

$$\Gamma^{(0)} = \Gamma_{2,0}^{(0)} P_{2,0},$$
where
$$\Gamma_{2,0}^{(0)} = \frac{5}{2} \frac{k(1-\beta)}{(25\lambda - 25\lambda\beta + 5k\beta)\delta + k\beta + 5(1+\lambda)(1-\beta)}. \tag{18}$$

Substituting the leading order solution for flow field and surfactant concentration into the leading order normal stress balance (that can be derived from equation (16)), the $O(Ca)$ correction to the droplet shape is obtained where the constant coefficients are as follows

$$L_{n,m}^{(Ca)} = \begin{cases} \dfrac{\left[80k\beta + 400\lambda(1-\beta)\right]\delta + 20k\beta + 5(16+19\lambda)(1-\beta)}{(200\lambda - 200\lambda\beta + 40k\beta)\delta + 8k\beta + 40(1+\lambda)(1-\beta)} & \text{for } m=0, n=2 \\ 0 & \text{otherwise.} \end{cases} \tag{19}$$

$$\hat{L}_{n,m}^{(Ca)} = 0.$$

The $O(Ca)$ surfactant concentration $\left[\Gamma^{(Ca)}\right]$ which is obtained by simultaneously solving the surfactant transport equation and the flow field boundary conditions of the same order can be expressed as follows

$$\Gamma^{(Ca)} = \Gamma^{(Ca)}_{2,0} P_{2,0} + \Gamma^{(Ca)}_{4,0} P_{4,0},$$

where

$$\Gamma^{(Ca)}_{2,0} = \frac{25\left(\alpha^{(1)}_{2,0}\delta^2 + \alpha^{(2)}_{2,0}\delta + \alpha^{(3)}_{2,0}\right)}{112\left[\alpha^{(4)}_{2,0}\delta + \alpha^{(5)}_{2,0}\right]^3},$$ (20)

$$\Gamma^{(Ca)}_{2,0} = \frac{45\left(\alpha^{(1)}_{4,0}\delta^2 + \alpha^{(2)}_{4,0}\delta + \alpha^{(3)}_{4,0}\right)}{112\left[\left\{\alpha^{(4)}_{4,0}\delta + \alpha^{(5)}_{4,0}\right\}^2 \left\{\alpha^{(6)}_{4,0}\delta + \alpha^{(7)}_{4,0}\right\}\right]}.$$

where the expression for the constants $\alpha^{(i)}_{j,0} \equiv \alpha^{(i)}_{j,0}(k,\beta,\lambda)$ $\left(i\in[1,5]\text{ for } j=2, i\in[1,7]\text{ for } j=4\right)$ are provided in Appendix A. The $O(Ca^2)$ shape correction similarly obtained from $O(Ca)$ normal stress balance is provided below

$$g^{(Ca^2)} = L^{(Ca^2)}_{2,0} P_{2,0} + L^{(Ca^2)}_{4,0} P_{4,0},$$

where (21)

$$L^{(Ca^2)}_{2,0} = \frac{25\left(\omega^{(1)}_{2,0}\delta^3 + \omega^{(2)}_{2,0}\delta^2 + \omega^{(3)}_{2,0}\delta + \omega^{(4)}_{2,0}\right)}{448\left[\omega^{(5)}_{2,0}\delta + \omega^{(6)}_{2,0}\right]^3}, \quad L^{(Ca^2)}_{4,0} = \frac{15\left(\omega^{(1)}_{4,0}\delta^3 + \omega^{(2)}_{4,0}\delta^2 + \omega^{(3)}_{4,0}\delta + \omega^{(4)}_{4,0}\right)}{224\left[\omega^{(5)}_{4,0}\delta + \omega^{(6)}_{4,0}\right]^2 \left[\omega^{(7)}_{4,0}\delta + \omega^{(8)}_{4,0}\right]},$$

where the constants, $\omega^{(i)}_{j,0} \equiv \omega^{(i)}_{j,0}(k,\beta,\lambda)$ $\left(i\in[1,6]\text{ for } j=2, i\in[1,8]\text{ for } j=4\right)$ in the above equation are presented in Appendix A.

The deformed shape of the droplet can thus be expressed as

$$r_s = 1 + Ca\left\{L^{(Ca)}_{2,0} P_{2,0}\right\} + Ca^2\left\{L^{(Ca^2)}_{0,0} + L^{(Ca^2)}_{2,0} P_{2,0} + L^{(Ca^2)}_{4,0} P_{4,0}\right\},$$ (22)

where $L^{(Ca^2)}_{0,0}$ is present to take care of the volume conservation constraint, that is mathematically given by [14]

$$\int_{\varphi=0}^{2\pi}\int_{\theta=0}^{\pi}\int_{r=0}^{r_s} r^2 dr \sin\theta d\theta d\varphi = \frac{4\pi}{3}.$$ (23)

Using the above constraint, it can be found out that equation (22) is satisfied only when

$$L_{0,0}^{(Ca^2)} = -\frac{1}{5}\left[L_{2,0}^{(Ca)}\right]^2. \tag{24}$$

For the limiting scenario of surface-convection-dominated surfactant transport ($k \to \infty$), the different constant coefficients present in the expression of the droplet shape (equation (22)) are given by

$$\lim_{k \to \infty} L_{2,0}^{(Ca)} = \frac{5}{2}\left(\frac{4\delta + 1}{1 + 5\delta}\right),$$

$$\lim_{k \to \infty} L_{2,0}^{(Ca^2)} = \frac{25}{28}\left[\frac{256\delta^3 + 244\delta^2 + 69\delta + 6}{(1+5\delta)^3}\right], \quad \lim_{k \to \infty} L_{4,0}^{(Ca^2)} = \frac{15}{28}\left[\frac{1312\delta^3 + 792\delta^2 + 152\delta + 9}{(1+5\delta)^2(1+9\delta)}\right]. \tag{25}$$

As can be seen from the above expression, interfacial slip has quite an impact on the droplet deformation in the high surface Péclet number limit although the same is independent of the surfactant distribution. The droplet deformation can be quantified in terms of a deformation parameter, $D_{fe}$, which for a uniaxial-extensional flow can be expressed as [9,14]

$$D_{fe} = \frac{r_s(\theta = 0) - r_s(\theta = \pi/2)}{r_s(\theta = 0) + r_s(\theta = \pi/2)}, \tag{26}$$

where the above expression when expanded in a binomial series takes the following form [14]

$$D_{fe} = \left\{\frac{3}{4}L_{2,0}^{(Ca)}\right\}Ca + \left[\frac{5}{16}L_{4,0}^{(Ca^2)} + \frac{3}{4}L_{2,0}^{(Ca^2)} - \frac{3}{16}\left\{L_{2,0}^{(Ca^2)}\right\}^2\right]Ca^2. \tag{27}$$

## B. Simple shear flow

For the case of an imposed shear flow, both the shape deformation as well as the surfactant concentration are obtained using a similar approach. Hence only the important results are highlighted below.

The leading order $O(Ca)$ surfactant concentration obtained by simultaneously solving the flow field boundary conditions and surfactant transport equation of the respective orders of perturbation are provided below

$$\Gamma^{(0)} = \hat{\Gamma}_{2,0}^{(0)} P_{2,0},$$
where
$$\hat{\Gamma}_{2,0}^{(0)} = \frac{5}{12}\frac{(1-\beta)k}{\{25\lambda(1-\beta) + 5k\beta\}\delta + k\beta + 5(1+\lambda)(1-\beta)}. \tag{28}$$

and

$$\Gamma^{(Ca)} = \Gamma^{(Ca)}_{2,0} P_{2,0} + \Gamma^{(Ca)}_{2,2} \cos 2\varphi P_{2,2} + \Gamma^{(Ca)}_{2,0} P_{2,0} + \Gamma^{(Ca)}_{4,4} \cos 4\varphi P_{4,4},$$

where

$$\Gamma^{(Ca)}_{2,0} = \frac{-25 \left[ \zeta^{(1)}_{2,0} \delta^2 + \zeta^{(2)}_{2,0} \delta + \zeta^{(3)}_{2,0} \right]}{336 \left[ \zeta^{(4)}_{2,0} \delta + \zeta^{(5)}_{2,0} \right]^3}, \quad \Gamma^{(Ca)}_{4,0} = \frac{5 \left\{ \zeta^{(1)}_{4,0} \delta^2 + \zeta^{(2)}_{4,0} \delta + \zeta^{(3)}_{4,0} \right\}}{224 \left[ \zeta^{(4)}_{4,0} \delta + \zeta^{(5)}_{4,0} \right]^2 \left[ \zeta^{(6)}_{4,0} \delta + \zeta^{(7)}_{4,0} \right]}, \quad (29)$$

$$\Gamma^{(Ca)}_{2,2} = \frac{5 \left[ \zeta^{(1)}_{2,2} \delta^2 + \zeta^{(2)}_{2,2} \delta + \zeta^{(3)}_{2,2} \right]}{288 \left[ \zeta^{(4)}_{2,2} \delta + \zeta^{(5)}_{2,2} \right]^2}, \quad \Gamma^{(Ca)}_{4,4} = \frac{-5 \left\{ \zeta^{(1)}_{4,4} \delta^2 + \zeta^{(2)}_{4,4} \delta + \zeta^{(3)}_{4,4} \right\}}{5376 \left[ \zeta^{(4)}_{4,4} \delta + \zeta^{(5)}_{4,4} \right]^2 \left[ \zeta^{(6)}_{4,4} \delta + \zeta^{(7)}_{4,4} \right]}.$$

The expressions of the constant coefficients, $\zeta^{(1)}_{2,0} - \zeta^{(5)}_{2,0}$, $\zeta^{(1)}_{2,2} - \zeta^{(5)}_{2,2}$, $\zeta^{(1)}_{4,0} - \zeta^{(7)}_{4,0}$ and $\zeta^{(1)}_{4,4} - \zeta^{(7)}_{4,4}$, present in equation (29) are given in Appendix B. The deformed shape of a droplet suspended in a simple shear flow can be represented by

$$r_s = \begin{bmatrix} 1 + Ca \left\{ L^{(Ca)}_{2,2} P_{2,2} \right\} \sin 2\varphi \\ + Ca^2 \left\{ L^{(Ca^2)}_{0,0} + L^{(Ca^2)}_{2,0} P_{2,0} + L^{(Ca^2)}_{2,2} P_{2,2} \cos 2\varphi + L^{(Ca^2)}_{4,0} P_{4,0} + L^{(Ca^2)}_{4,4} P_{4,4} \cos 4\varphi \right\} \end{bmatrix}, \quad (30)$$

where the constant coefficients in the above equation are provided below

$$\hat{L}^{(Ca)}_{2,2} = \frac{\left[ 400 \lambda (1-\beta) + 80 k\beta \right] \delta + 5(16+19\lambda)(1-\beta) + 20 k\beta}{\left[ 1200 \lambda (1-\beta) + 240 k\beta \right] \delta + 240(1+\lambda)(1-\beta) + 48 k\beta}, \quad (31)$$

and

$$\left. \begin{array}{c} L^{(Ca^2)}_{2,0} = \dfrac{\xi^{(1)}_{2,0} \delta^3 + \xi^{(2)}_{2,0} \delta^2 + \xi^{(3)}_{2,0} \delta + \xi^{(4)}_{2,0}}{1344 \left\{ \xi^{(5)}_{2,0} \delta + \xi^{(6)}_{2,0} \right\}^3}, \quad L^{(Ca^2)}_{2,2} = \dfrac{\xi^{(1)}_{2,2} \delta^2 + \xi^{(2)}_{2,2} \delta + \xi^{(3)}_{2,2}}{1152 \left\{ \xi^{(4)}_{2,2} \delta + \xi^{(5)}_{2,2} \right\}^2}, \\[2ex] L^{(Ca^2)}_{0,4} = \dfrac{\xi^{(1)}_{4,0} \delta^3 + \xi^{(2)}_{4,0} \delta^2 + \xi^{(3)}_{4,0} \delta + \xi^{(4)}_{4,0}}{1344 \left\{ \xi^{(5)}_{4,0} \delta + \xi^{(6)}_{4,0} \right\}^2 \left\{ \xi^{(7)}_{2,2} \delta + \xi^{(8)}_{2,2} \right\}}, \quad L^{(Ca^2)}_{4,4} = \dfrac{\xi^{(1)}_{4,4} \delta^3 + \xi^{(2)}_{4,4} \delta^2 + \xi^{(3)}_{4,4} \delta + \xi^{(4)}_{4,4}}{32256 \left[ \xi^{(5)}_{4,4} \delta + \xi^{(6)}_{4,4} \right]^2 \left\{ \xi^{(7)}_{2,2} \delta + \xi^{(8)}_{2,2} \right\}}, \end{array} \right\} \quad (32)$$

where the expressions of the constants, $\xi^{(1)}_{2,0} - \xi^{(6)}_{2,0}$, $\xi^{(1)}_{2,2} - \xi^{(5)}_{2,2}$, $\xi^{(1)}_{4,0} - \xi^{(8)}_{4,0}$ and $\xi^{(1)}_{4,4} - \xi^{(8)}_{4,4}$, present in (32) are given in Appendix B. The term $L^{(Ca^2)}_{0,0}$ in equation (30) takes care of the volume conservation constraint [equation (23)] and is given by

$$L^{(Ca^2)}_{0,0} = -\frac{12}{5} \left\{ \hat{L}^{(Ca)}_{2,2} \right\}^2. \quad (33)$$

In the limiting case of surface-convection dominated surfactant transport $(k \to \infty)$, the respective constants present in the expression of the droplet shape [equation (30)] are given by

$$\lim_{k \to \infty} \hat{L}_{2,2}^{(Ca)} = \frac{5}{12}\left(\frac{4\delta+1}{5\delta+1}\right), \quad \lim_{k \to \infty} L_{2,0}^{(Ca^2)} = -\frac{25}{84}\left\{\frac{244\delta^2 + 69\delta + 256\delta^3 + 6}{(5\delta+1)^3}\right\},$$

$$\lim_{k \to \infty} L_{2,2}^{(Ca^2)} = \frac{5}{288}\left[\frac{2\{(230\lambda+288)\delta^2+(101\lambda+136)\delta+11\lambda+14\}}{(5\delta+1)^2} + \frac{\lambda+5\delta\lambda+4}{\beta(5\delta+1)^2}\right], \quad (34)$$

$$\lim_{k \to \infty} L_{4,0}^{(Ca^2)} = \frac{5}{168}\left\{\frac{1312\delta^3 + 792\delta^2 + 152\delta + 9}{(5\delta+1)^2(1+9\delta)}\right\}, \quad \lim_{k \to \infty} L_{4,4}^{(Ca^2)} = -\frac{5}{4032}\left\{\frac{1312\delta^3 + 792\delta^2 + 152\delta + 9}{(5\delta+1)^2(1+9\delta)}\right\}.$$

The deformation parameter for the case of an imposed shear flow used to represent the magnitude of the droplet deformation along the shear plane $(\theta = \pi/2)$ can be expressed as [9]

$$D_{fs} = \frac{\max\left[r_s(\theta=\pi/2,\varphi)\right] - \min\left[r_s(\theta=\pi/2,\varphi)\right]}{\max\left[r_s(\theta=\pi/2,\varphi)\right] + \min\left[r_s(\theta=\pi/2,\varphi)\right]}. \quad (35)$$

The inclination angle $(\varphi_d)$ of a droplet suspended in a shear flow is also a parameter of interest, that has been previously investigated in several studies[12–14]. The magnitude of the inclination angle for any particular value of $\theta$ is equal to $\varphi$ corresponding to the maximum value of $r_s$.

## IV. SUSPENSION RHEOLOGY

We now investigate how interfacial slip is responsible for altering the effective viscosity of a dilute suspension of surfactant-laden droplets in either type of linear flows considered. A dilute emulsion of droplets can be expressed as $\phi \ll 1$, where $\phi$ is the volume fraction of the droplet phase. With the leading order and $O(Ca)$ flow fields at hand, we aim at finding out the $O(\phi)$ and $O(\phi Ca)$ correction in the effective viscosity of the suspension. Following Batchelor, the volume averaged suspension stress of a dilute emulsion of force-free particles can be expressed as [9,37]

$$\langle \tau_s \rangle = -\langle p \rangle \mathbf{I} + 2\mathbf{D}_\infty + \frac{\phi}{V_d}\mathbf{S}, \quad (36)$$

where $\mathbf{D}_\infty$ has been defined previously in equations (2) and (3) and $\mathbf{S}$ is a symmetric stresslet that indicates the change in total stress induced due to the presence of a droplet in the flow field. The stress can be mathematically represented as [9]

$$\mathbf{S} = \int_{\varphi=0}^{2\pi} \int_{\theta=0}^{\pi} \left[ \frac{1}{2}\{(\boldsymbol{\tau}\cdot\mathbf{n})\mathbf{x} + ((\boldsymbol{\tau}\cdot\mathbf{n})\mathbf{x})^T\} - \frac{1}{3}\mathbf{I}\{(\boldsymbol{\tau}\cdot\mathbf{n})\cdot\mathbf{x}\} - \{\mathbf{un} + (\mathbf{un})^T\} \right] r_s^2 d\theta d\varphi. \qquad (37)$$

The effective extensional viscosity or the Trouton viscosity of a dilute emulsion of surfactant-laden droplets suspended in a uniaxial extensional flow is given by

$$\frac{\mu_{ext}}{\mu_e} = \frac{\langle\overline{\tau}_{zz}\rangle - \langle\overline{\tau}_{yy}\rangle}{\mu_e \dot{\gamma}} = \frac{\langle\overline{\tau}_{zz}\rangle - \langle\overline{\tau}_{xx}\rangle}{\mu_e \dot{\gamma}}$$
$$3\left[1 + \frac{5}{2}\left\{\frac{\varepsilon_1\delta + \varepsilon_2}{\varepsilon_3\delta + \varepsilon_4} + \frac{15(\varepsilon_5\delta^3 + \varepsilon_6\delta^2 + \varepsilon_7\delta + \varepsilon_8)}{56(\varepsilon_3\delta + \varepsilon_4)^3} Ca\right\}\phi\right], \qquad (38)$$

where all the unknown constants are provided in Appendix C. It should be noted that for a surfactant-free droplet or on substitution of $\beta = 0$ in equation (38), we obtain exactly the same expression for non-dimensional Trouton viscosity as was derived by Ramachandran and Leal in their study [9]. The effective extensional viscosity for the limiting scenario of surface convection-dominated surfactant transport is provided below

$$\lim_{k\to\infty} \frac{\mu_{ext}}{\mu_e} = 3\left[1 + \frac{5}{28}\left\{\frac{14(2\delta+1)}{(5\delta+1)} + \frac{15(4\delta+1)(4\delta^2+8\delta+1)}{(5\delta+1)^3}Ca\right\}\phi + O(Ca^2)\right] \qquad (39)$$

For the case of imposed simple shear flow, the effective viscosity of the dilute suspension, for the limiting scenario of $Pe_s \ll 1$, is expressed as

$$\frac{\mu_{eff}}{\mu_e} = \frac{\langle\overline{\tau}_{xy}\rangle}{\mu_e \dot{\gamma}} = 1 + \frac{5}{2}\left[\frac{\{2k\beta + 10\lambda(1-\beta)\}\delta + (1-\beta)(2+5\lambda) + k\beta}{\{25\lambda(1-\beta) + 5k\beta\}\delta + 5(1+\lambda)(1-\beta) + k\beta}\right]\phi + O(Ca^2). \qquad (40)$$

From the above expression, it can be seen that there is no effect of shape deformation on the effective viscosity till $O(Ca)$. The first and second normal stress differences $(N_1$ and $N_2)$ are next provided below

$$N_1 = \langle \bar{\tau}_{xx} \rangle - \langle \bar{\tau}_{yy} \rangle = \frac{5\left\{\varpi_1^{(1)}\delta^2 + \varpi_1^{(2)}\delta + \varpi_1^{(3)}\right\}}{8\left\{\varpi_1^{(4)}\delta + \varpi_1^{(5)}\right\}^2}\phi Ca,$$

$$N_2 = \langle \bar{\tau}_{yy} \rangle - \langle \bar{\tau}_{zz} \rangle = -\frac{5\left(\varpi_2^{(1)}\delta^3 + \varpi_2^{(2)}\delta^2 + \varpi_2^{(3)}\delta + \varpi_1^{(4)}\right)}{112\left\{\varpi_2^{(5)}\delta + \varpi_2^{(6)}\right\}^3}\phi Ca.$$

(41)

where unknown constants, $\varpi_j^{(i)}$ $\left(i \in [1,5] \text{ for } j=1, i \in [1,6] \text{ for } j=2\right)$ appearing in equation (41) are given in Appendix C. Again, in the limiting scenario of $k \to \infty$, the effective viscosities and the normal stress differences are obtained as

$$\lim_{k \to \infty}\frac{\mu_{eff}}{\mu_e} = 1 + \frac{5}{2}\left(\frac{2\delta+1}{5\delta+1}\right)\phi, \lim_{k \to \infty} N_1 = \frac{5}{2}\left\{\frac{64\delta^2\beta + 32\delta\beta + 3\beta + 1}{\beta(5\delta+1)^2}\right\}\phi Ca,$$

$$\lim_{k \to \infty} N_2 = -\frac{5}{28}\left\{\frac{2000\delta^3\beta + 1028\delta^2\beta + (149\beta+35)\delta + 6\beta + 7}{\beta(5\delta+1)^3}\right\}\phi Ca.$$

(42)

## V. RESULTS AND DISCUSSIONS

### A. Droplet deformation

#### 1. Uniaxial extensional flow

The theoretical results, thus obtained for the limiting case of low surface Péclet number, are first validated against the results from the experimental study performed by Hu and Lips [20]. In this study, Hu and Lips investigated the effect of surfactant distribution along the interface of the droplet deformation. They used Polydimethylsiloxane as the suspending phase and Polybutadiene as the dispersed phase. In fig 2, we have compared our theoretical prediction with the experimental results reported in fig 3(a) of their paper. They have provided the experimental data corresponding to different values of surfactant coverage on the droplet surface. In their study, Hu and Lips have not taken into account the effect of interfacial slip on droplet deformation. On the contrary, the deformation parameter, defined for uniaxial extensional flow in equation (27), is plotted as a function of $Ca$ corresponding to a finite value of the slip factor $(\delta = 0.5)$ in fig. 2. As there is no specific mention of the values of $k$ or $\beta$ in the work by Hu and Lips, we take the freedom to choose typical values of $k$ and $\beta$ for commercially available surfactants, that is $\beta = 0.5$, and $k = 2$. The deformation parameter based on both $O(Ca^2)$ and $O(Ca)$ shape corrections of the droplet are plotted. It can be seen from fig. 2 that there is a pretty good match between the $O(Ca^2)$ theoretical prediction and the experimental results of Hu and Lips[20] in the low capillary number regime $(Ca \leq 0.1)$ whereas for higher values of $Ca$, the analytical results are seen to deviate from the experimental results.

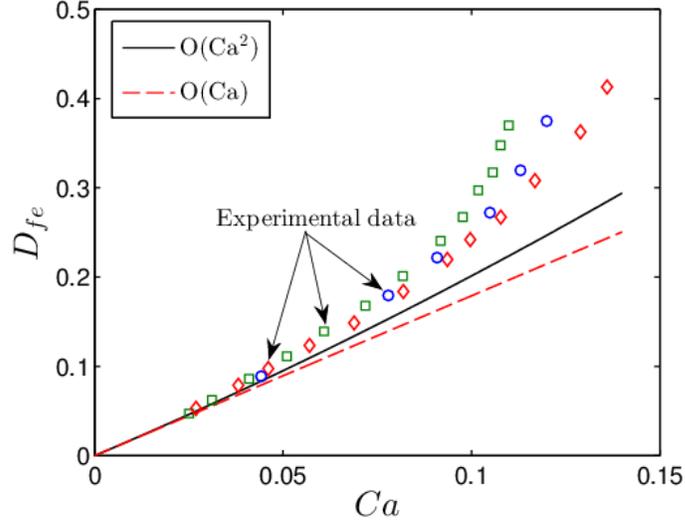

Fig. 2: Variation of $O(Ca)$ and $O(Ca^2)$ deformation parameter with $Ca$, when the droplet is suspended in a uniaxial extensional flow. The above plot is shown for the limiting case of $Pe_s \ll 1$. The marker points in the plot denote the experimental data for different surfactant coverage as extracted from fig. 3(a) of Hu and Lips [20]. The different parameter values taken for the above plot are $\lambda = 2.3$, $\beta = 0.5$, $k = 2$, and $\delta = 0.5$.

In order to show the effect of interfacial slip on droplet deformation, we plot deformation parameter $(D_{fe})$ as a function of the dimensionless slip factor for different values of $\beta$ and $k$ in fig. 3. In either of the plots it can be observed, that irrespective of any value of $\beta$ or $k$, the droplet deformation always reduces with increase in the interfacial slip. Fig. 3(a) shows the variation of $D_{fe}$ as a function of $\delta$ (under the limiting case of low $Pe_s$) for different values of the elasticity parameter, $\beta$ $(= 0.1, 0.4, 0.8)$. Although there is a monotonous decrease in droplet deformation with rise in interfacial slip, the rate of reduction of droplet deformation or the influence of slip on the same becomes increasingly insignificant for lower values of $\beta$. Another important observation that can be made from fig. 3(a) is that higher values of $\delta$ render any influence of $\beta$ on droplet deformation, ineffective. That is, for larger interfacial slip, rise in droplet deformation due to increase in $\beta$ is relatively low as compared to the increase for low values of $\delta$.

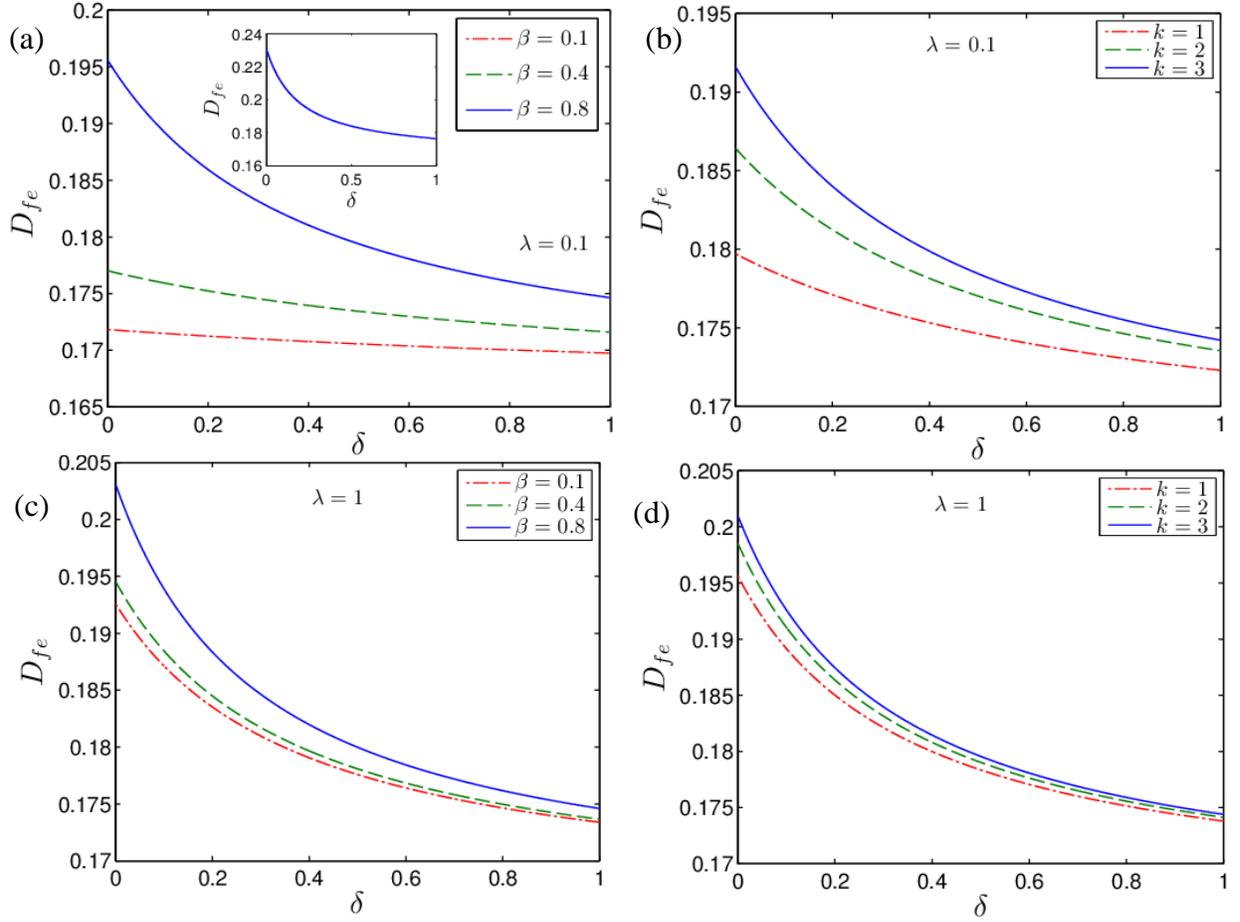

Fig. 3: Variation of $O(Ca^2)$ deformation parameter with $\delta$ for different values of $k$ and $\beta$ (under the limiting case of low $Pe_s$), when the droplet is suspended in a uniaxial extensional flow. The inset in fig. (a) shows the corresponding variation for the high $Pe_s$ limit. In fig. (a) $\lambda = 0.1$, $k = 1$ and $\beta = 0.1, 0.4, 0.8$, in fig. (b) $\lambda = 0.1$, $\beta = 0.5$ and $k = 1, 2, 3$, in fig. (c) $\lambda = 1$, $k = 1$ and $\beta = 0.1, 0.4, 0.8$ and in fig. (d) $\lambda = 1$, $\beta = 0.5$ and $k = 1, 2, 3$. The value of $Ca$ is taken as 0.1.

Figure 3(b) shows the variation of the droplet deformation as function of $\delta$ for different values of the property parameter $(k = 1, 2, 3)$. It can be seen that for any particular value of $\delta$, the droplet deformation is enhanced with an increase in $k$, which is similar to the behavior as reported by Mandal et al. [14]. Analogical to the previous case (fig. 3(a)), the effect of $k$ gradually becomes insignificant with increase in $\delta$. The role played by droplet viscosity or the viscosity ratio $(\lambda)$ on droplet deformation can be understood on comparison of figures 3(a) and 3(c) as well as figures 3(b) and 3(d). In both figures 3(c) and 3(d), the variation of $D_{fe}$ is shown as a function of $\delta$ for different values of $\beta$ and $k$, respectively, for the particular case of $\lambda = 1$ as compared to figures 3(a) and 3(b) where $\lambda = 0.1$. It can be inferred from figure 3 that increase in

$\lambda$ is always accompanied by an increase in the rate of decrease of droplet deformation with rise in the slip factor, $\delta$. In other words, higher values of $\lambda$ ensures higher effectiveness of interfacial slip towards reducing droplet deformation.

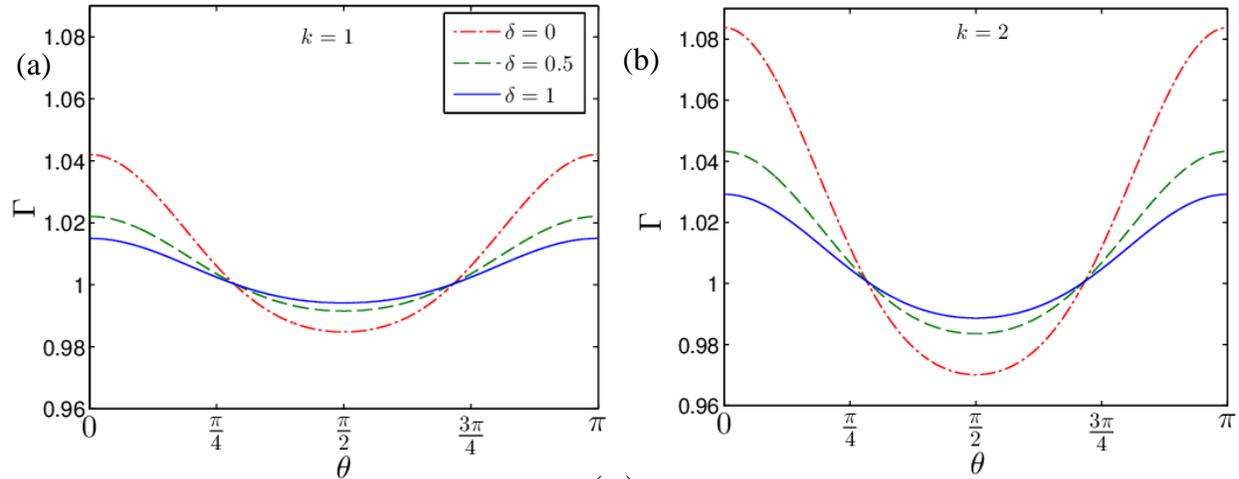

Fig. 4: Variation of surfactant concentration $(\Gamma)$ along the droplet surface for different values of $\delta$ under the limiting scenario of low $Pe_s$. Fig. (a) is plotted for $k=1$ whereas for fig. (b) we take $k=2$. The other parameter values used for the plot are $\lambda = 0.5$, $\beta = 0.1$, and $Ca = 0.1$.

To provide a physical explanation regarding the nature of variation of droplet deformation, we look into how the distribution of surfactants along the droplet surface is affected by interfacial slip. It should be noted that the presence of slip results in a discontinuity in the interfacial fluid flow velocity across the droplet surface. The fluid flow inside the droplet thus reduces in comparison to that outside the droplet which in turn results in a fall in the flow strength along the interface. Thus the surfactant transport along the droplet surface gets hampered due to the presence of interfacial slip. In other words, the asymmetry in surfactant distribution along the droplet surface as a result of the imposed linear flow is reduced due to the presence of interfacial slip. This behavior can be seen in figures 4(a) and 4(b), where the effect of $\delta$ on the surfactant distribution along the droplet surface is shown for $k=1$ and $k=2$ respectively. It can be seen that the presence of imposed uniaxial extensional flow results in a lower concentration of surfactants at the equatorial position of the droplet whereas there is a high concentration of surfactants at the poles. However, as discussed earlier, the imposed flow induced asymmetry in surfactant concentration is opposed by the interfacial slip which reduces any surfactant transport. Thus increase in interfacial slip decreases the gradient in surfactant concentration $\left(\left|\Gamma_{max} - \Gamma_{min}\right|\right)$ which is accompanied by a fall in the surface tension gradient along the droplet surface. This signifies a reduction in the net Marangoni stress, responsible for droplet deformation. The above discussion provides a clear explanation regarding the decrease in droplet deformation with rise in $\delta$. On further increase in $\delta$, the interfacial flow strength reduces, and any rise in droplet deformation due to increase in $k$ or $\beta$ becomes gradually insignificant (fig. 3). On comparison of fig. 4(a) with 4(b), it can be seen that for any particular value of $\delta$, $|\Gamma_{max} - \Gamma_{min}|$ increases with rise in $k$ which indicates a corresponding increase in the Marangoni stress. Thus for a constant $\delta$, droplet deformation increases with $k$. A similar behavior is also seen for the case increase in $\beta$. This can be observed from fig. 3 as well. Finally, for a higher droplet viscosity as compared to the carrier phase, the interfacial tension gradually becomes independent of the surfactant distribution. This, in turn, decreases the Marangoni stress, which accompanied by the presence of interfacial slip results in a higher rate of reduction in droplet deformation, as seen from figures 3(a) and 3(c)

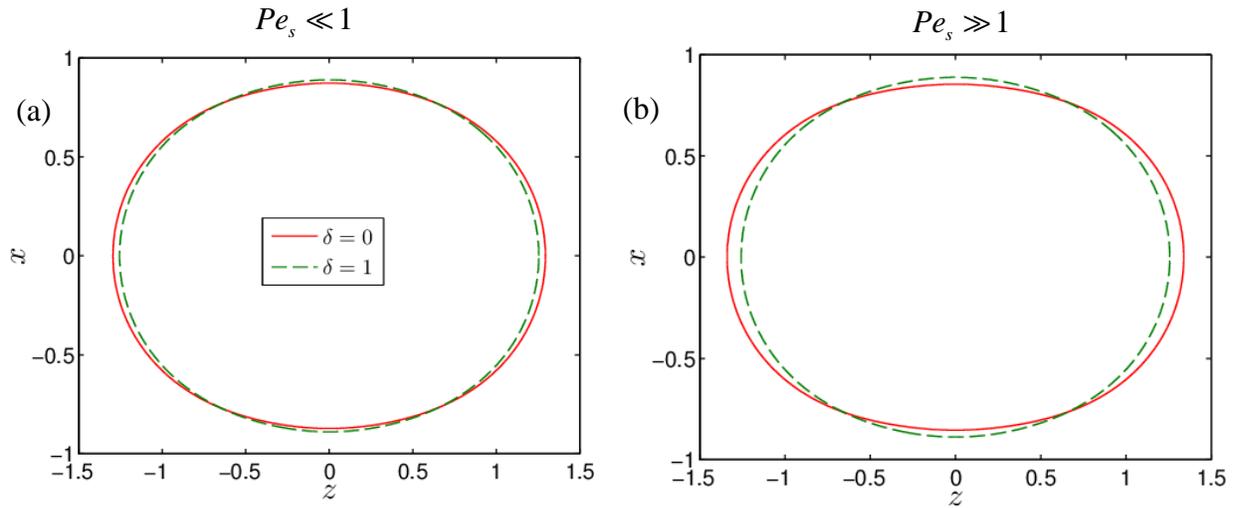

Fig. 5: Deformed shape of the droplet with and without the presence of interfacial slip $(\delta = 0,1)$. Fig. (a) is plotted for the limiting case of $Pe_s \ll 1$ $(k=2)$, whereas fig. (b) is for the limiting case of $Pe_s \gg 1$ $(k \to \infty)$. The parameter values used for the plot are $\lambda = 0.5$, $\beta = 0.7$ and $Ca = 0.1$.

For the limiting case of high $Pe_s$, surfactant transport along the droplet surface is mainly due to interfacial convection rather than diffusion. Hence, under this limit, the Marangoni stress developed is larger ensuring a larger droplet deformation. The effect of interfacial slip on droplet deformation is also higher in this limiting scenario, as surface convection is the sole mode of surfactant transport. It can be inferred from the expressions of the constants in equation (25) that droplet deformation is solely dependent on interfacial slip. Thus for the same increase in $\delta$, there is a larger decrease in droplet deformation for this limit as compared to the case of $Pe_s \ll 1$. This is evident from the inset in fig. 3(a).

We, now, take a look at the droplet shape with and without the presence of interfacial slip for both the limiting cases based on the mode of surfactant transport. Figure 5(a) shows the shape of the droplet in the presence an imposed uniaxial extensional flow for $\delta = 0$ and 1 under the limiting case of low $Pe_s$. Figure 5(b), on the other hand, shows the deformed droplet shape for the limiting scenario of high $Pe_s$. The droplet, in the presence of the uniaxial extensional flow, is found to get elongated into a prolate shape with the major axis along with the axis of extension. Clearly from fig. 5(a) as well as 5(b), presence of interfacial slip results in a reduced deviation from the original spherical shape of the droplet. On comparison of both fig. 5(a) and 5(b), it can be stated that for the same rise in interfacial slip, there is a lower droplet deformation for the limiting case of $Pe_s \ll 1$.

## 2. Simple shear flow

We, now, discuss the results obtained for the case when the droplet is suspended in a simple shear flow. Towards validating our theoretical prediction for the limiting case of low surface Péclet number, we compare our results with that obtained by Feigl et al. in their experimental study [23]. We show the variation of the shape parameters $L$, $B$ and $W$ as a function of $Ca$ in fig. 6 and compare them with the experimental data extracted from fig. 4 of the work done by Feigl et al. [23].

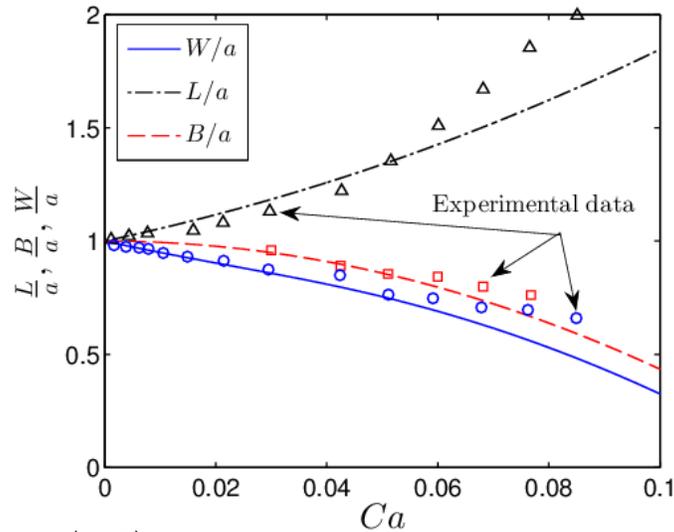

Fig. 6: Variation of $O(Ca^2)$ solution for the normalized deformation parameters: $L/a$, $B/a$ and $W/a$, with $Ca$. The lines indicate our theoretical prediction while circular, square and triangular marker points denote the experimental results for $W/a$, $B/a$ and $L/a$ respectively, as extracted from fig. 4 of Feigl. et. al.[23]. The values of the other parameters are $\delta = 0.5$, $\beta = 0.8$, $k = 1$ and $\lambda = 0.335$.

These parameters, which determine the magnitude of droplet deformation can be defined for the case of an imposed shear flow as

$$\begin{aligned} L &= \max_{\varphi}\{r_s(\theta = \pi/2, \varphi \in [0,\pi])\}, \\ B &= \min_{\varphi}\{r_s(\theta = \pi/2, \varphi \in [0,\pi])\}, \\ W &= \min_{\theta}\{r_s(\varphi = \pi/2, \theta \in [0,\pi/2])\}. \end{aligned} \qquad (43)$$

As can be seen from fig. 6, there is a pretty good match between our $O(Ca^2)$ theoretical prediction and the experimental results for all the parameters $(L, B, W)$.

We next put forward the variation of the droplet deformation parameter $(D_{fs})$ as a function of the interfacial slip factor, $\delta$ for different values of $k$ in fig. 7. The effect of change in $\beta$ on $D_{fs}$ follows the same trend as is shown by any variation in $k$. It is observed from fig. 7 that the droplet deformation reduces due to increase in $\delta$ for a constant value of $k$. Interfacial slip reduces the imposed shear flow induced interfacial fluid flow and hence the surfactant transport decreases thus reducing the gradient in surfactant concentration $(|\Gamma_{max} - \Gamma_{min}|)$ across the droplet surface. This is evitable from fig. 8.

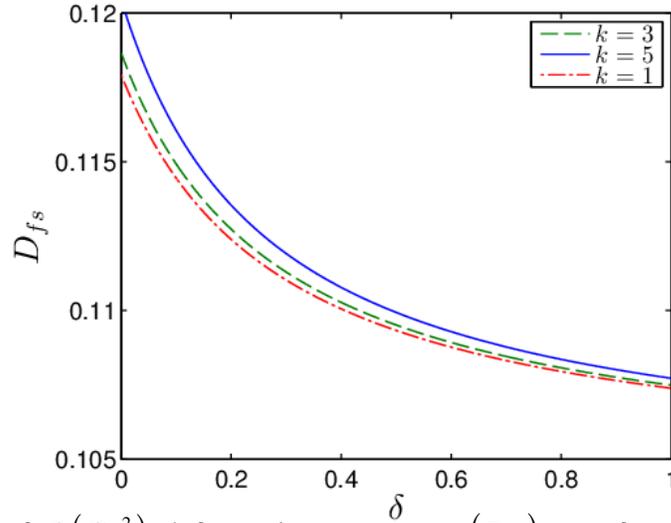

Fig. 7: Variation of $O(Ca^2)$ deformation parameter $(D_{fs})$ as a function of $\delta$, for a droplet suspended in a simple shear flow. The above plot is done for different values of $k (= 1, 3, 5)$ for the limiting case of low $Pe_s$. The values of the other parameters involved are $\beta = 0.5$, $\theta = \pi/2$, $Ca = 0.1$ and $\lambda = 2$.

Figure 8 shows the variation of the surfactant concentration along a transverse plane $(\varphi = 3\pi/4)$ for different values of the interfacial slip parameter. The surfactant induced Marangoni stress thus decreases with increase in interfacial slip. Hence a decrease in the droplet deformation with rise in interfacial slip is expected. For an imposed simple shear flow, however, the rise in deformation due to increase in either $\beta$ or $k$ is quite low for any constant value of $\delta$ (see fig. 7). The influence of viscosity ratio on droplet deformation in presence of interfacial slip remains the same as in the case of a uniaxial extensional flow and hence not highlighted here.

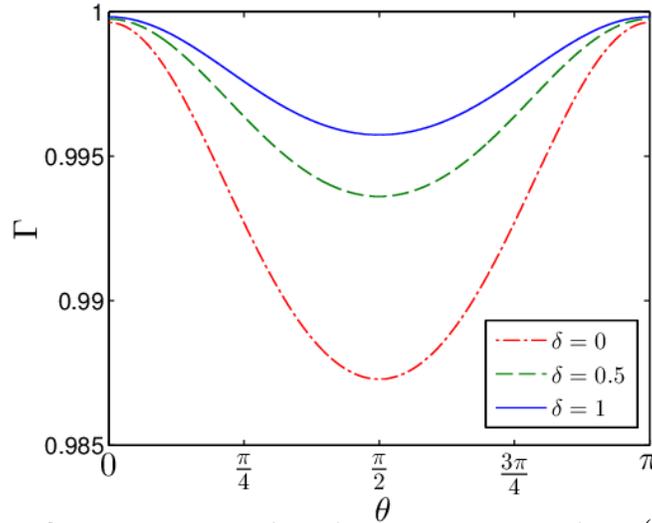

Fig. 8: Variation of surfactant concentration along a transverse plane $(\varphi = 3\pi/4)$ for different values of $\delta\ (=0, 0.5, 1)$. The above plot is shown for the case a droplet suspended in a simple shear flow under the limiting case of low $Pe_s$. The values of the other parameters involved are $k = 1$, $\beta = 0.5$, $Ca = 0.1$ and $\lambda = 0.5$.

We next look into the variation of the inclination angle and how it is affected by the presence of interfacial slip. The inclination angle, $\varphi_d$, for any particular value of $\theta$ can be obtained from the magnitude of $L$, which is the droplet dimension measured along the axis of applied shear. Figure 9 shows the variation of inclination angle as a function of $Ca$ with and without the presence of interfacial slip $(\delta = 0, 1)$. It is seen that presence of slip at the droplet surface, reduces the inclination angle. The effect of interfacial slip is however prominent at higher values of $Ca$. With increase in $Ca$, on the other hand, there is a monotonous fall in $\varphi_d$. We have also provided a comparison of our theoretical result for inclination angle with the data extracted from the experimental study by Feigl et al [23]. As seen, there is a better match between the theoretical and experimental results when interfacial slip is taken into consideration.

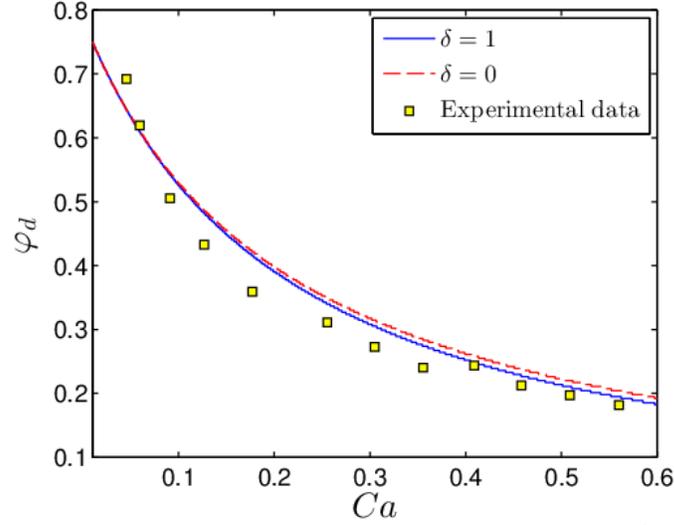

Fig. 9: Plot displaying the variation of the droplet inclination angle $(\varphi_d)$ as a function of $Ca$ for the case of low $Pe_s$. The square marker points denote the experimental data as extracted from the work of Feigl et al. [23]. The values of the different parameters involved are $k = 5$, $\beta = 0.5$ and $\lambda = 6.338$.

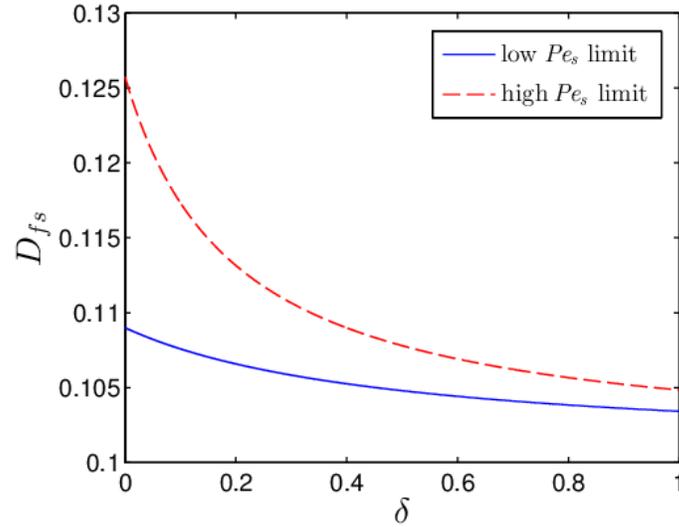

Fig. 10: A comparison of the variation of deformation parameter $(D_{fs})$ of a droplet (suspended in a simple shear flow) with $\delta$ between the two limiting case of high and low $Pe_s$. The values of the other parameters involved are $k = 1$ (for low Pes limit) $\beta = 0.5$, $\theta = \pi/2$, $Ca = 0.1$ and $\lambda = 0.5$.

In the limiting case of surface convection driven surfactant transport or the high surface Péclet number limit, the interfacial slip is highly effective in reducing the droplet deformation as compared to the other limiting case of low $Pe_s$. This is evident from fig. 10, where we have compared the decrease in droplet deformation between the two limiting cases of high and low $Pe_s$ for the same increase in $\delta$. The other different parameters used for this plot are provided in the figure caption.

The above observation can also be made from the deformed droplet shape under the two limiting cases as shown in fig. 11(a) and 11(b). For each of the limiting cases, the droplet shape is shown with and without the presence of any interfacial slip. Presence of slip in either case, prevents the droplet from deviating from its spherical shape. The droplet is seen to elongate along the axis of imposed shear and has a finite inclination angle that depends on the imposed shear rate as well as on the magnitude of the interfacial slip. For the limiting case of high $Pe_s$, the droplet inclination angle is seen to vary in almost a similar way as observed for the limiting case of low $Pe_s$. It can also be seen that droplet deformation is more severely affected due to the presence of slip for the case of an imposed uniaxial- extensional flow as compared to the case of simple shear flow.

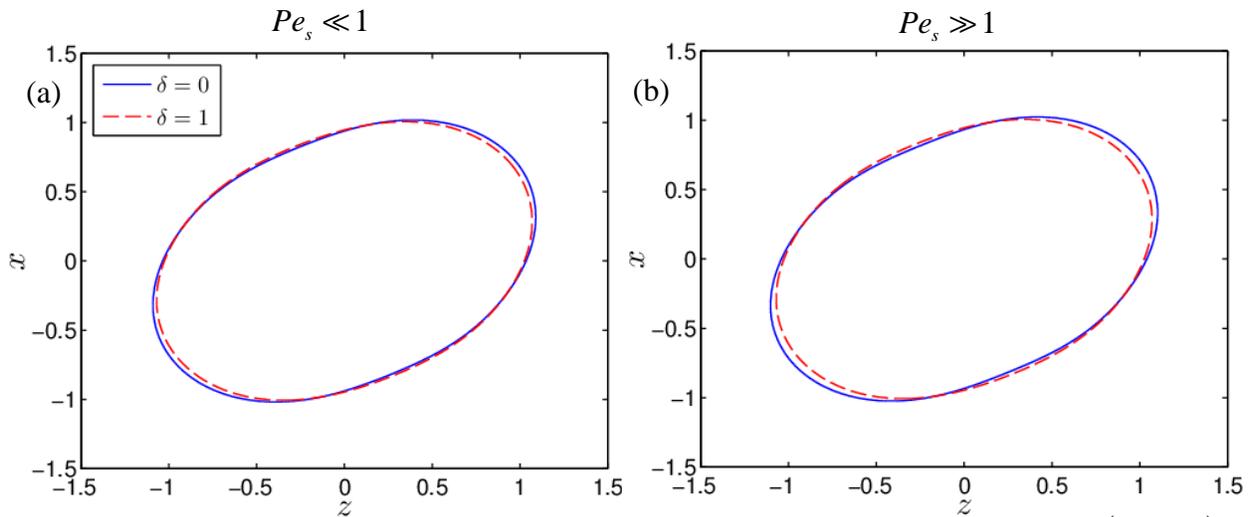

Fig. 11: Deformed droplet shape with and without the presence of interfacial slip $(\delta = 0,1)$. Fig. (a) is plotted for the limiting case of $Pe_s \ll 1$ $(k=2)$, whereas fig. (b) is for the limiting case of $Pe_s \gg 1$. The parameter values used for the plot are $\lambda = 0.5$, $\beta = 0.7$ and $Ca = 0.15$.

## B. Suspension Rheology

### 1. Uniaxial extensional flow

We now look into the effect of interfacial slip on the effective viscosity of a dilute emulsion of droplets in a uniaxial extensional flow. The expression for the dimensionless effective extensional viscosity is provided in equation (38) and is found to match exactly with the expression provided by Ramachandran and Leal [9] in their work for the special case of a surfactant-free droplet. Presence of particles or droplets present in any emulsion always enhances its effective extensional viscosity due to the resistance offered by them against the imposed flow [9]. It has been shown in previous studies that for a clean droplet, this resistance is enhanced with increase in droplet viscosity which results in a rise of effective viscosity [9,14]. For the case of a surfactant-laden droplet the surfactants get redistributed along the droplet surface due to the imposed flow and hence a Marangoni stress develops due to the variation of surface tension along the interface. This Marangoni stress further increases the resistance to the bulk flow and in turn enhances the effective extensional viscosity [14]. The presence of interfacial slip, however, has a negative effective on the effective viscosity or the Trouton viscosity as can be seen from fig. 12, where we have shown the variation of normalized Trouton viscosity $\left(\mu_{ext}/\mu_e - 3\right)/\phi$ with the dimensionless slip factor $(\delta)$ for different values of $k$. Presence of slip at the interface, reduces the fluid flow and hence the nonuniformity in surfactant distribution. The surfactant-induced Marangoni stress, as a result, thus reduces which thus decreases the overall resistance provided by the droplets. The effective viscosity, hence for any particular value of $k$, decreases which can be observed from fig. 12. Increase in $k$, for a constant value of slip, results in an increase in the surface convection of surfactants. Thus the Marangoni stress increases followed by a rise in the effective viscosity. However, such a rise is rendered ineffective at high values of $\delta$, where any increase in Marangoni stress, due to increase in $k\left(\text{or }\beta\right)$, is dominated by the reduction in interfacial fluid flow due to the presence of slip. A similar rise in the effective viscosity is also observed due to increase in $\beta$ and $\lambda$ for any constant value of $\delta$.

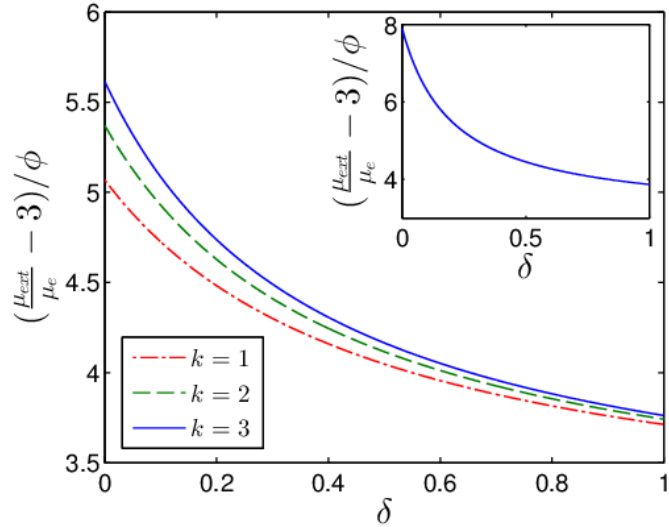

Fig. 12: Variation of normalized effective extensional viscosity with $\delta$ for different values of the property parameter, $k$. The inset in the above figure shows the variation for the limiting case of high $Pe_s$. The different parameter values used are $\lambda = 0.5$, $\beta = 0.75$ and $Ca = 0.05$.

For the limiting case of high surface Péclet number, surface convection is the main mode of surfactant transport along the interface and hence interfacial slip has a significantly larger effect on the effective viscosity. It can be observed from the inset of fig. 12 that under this limit, there is a greater fall in the effective viscosity for the same increase in the interfacial slip.

## 2. Simple shear flow

The expression for the non-dimensional form of effective viscosity of a dilute suspension of droplets in a shear flow is provided in equation (40). From the expression it is evident that it is independent of droplet deformation and at the same time exhibits a similar behavior due to variation in interfacial slip, as was observed for the case of a uniaxial extensional flow. However, deformation of a droplet in a shear flow generates normal stress differences, $N_1$ and $N_2$, which gives rise to a non-Newtonian behavior of the emulsion[38]. Towards highlighting the effect of interfacial slip on both $N_1$ and $N_2$, we plot the variation of the normalized normal stress differences $(N_1/\phi, N_2/\phi)$ as a function of $\delta$, for different values of $k$ in fig. 13(a) and 13(b) respectively.

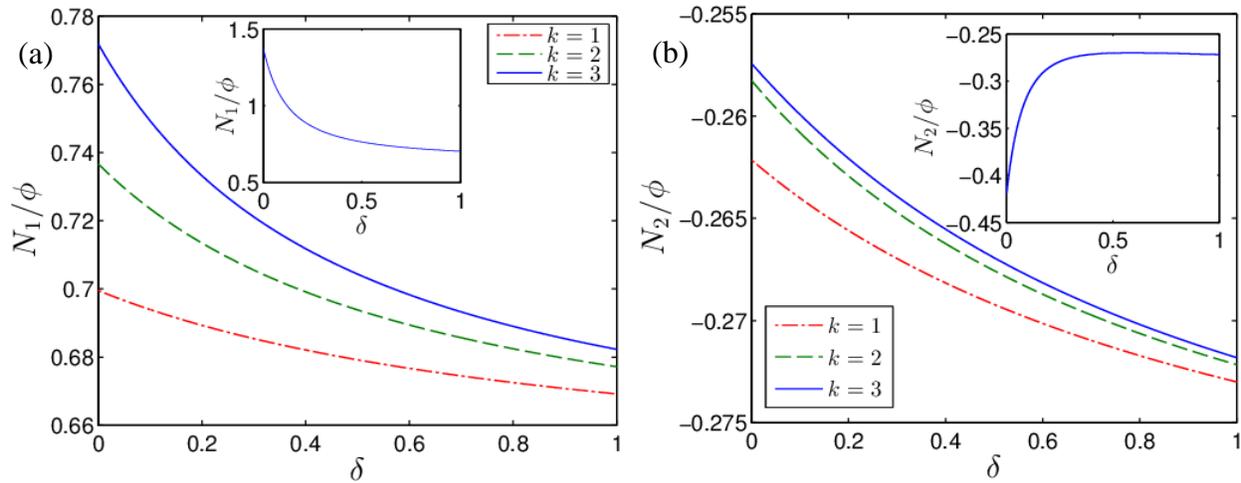

Fig. 13: Variation of the normalized first and second normal stress differences $(N_1/\phi, N_2/\phi)$ with $\delta$ for different values of $k$ $(=1, 2, 3)$. The inset in both fig. (a) and (b), shows the corresponding variation for the limit of high surface Péclet number. The other parameter values used are $\lambda = 0.1$, $\beta = 0.4$ and $Ca = 0.1$.

It can be seen from fig. 13(a) and 13(b) that $N_1$ increases while $N_2$ reduces due to increase in $k$ for a particular value of $\delta$. The positive sign of $N_1$ is due to the tensile component of the surface tension that acts along the axis of shear, whereas the negative sign of $N_2$ indicates compressive stresses that act on the droplet along the gradient direction.

On a careful analysis of fig. 13(a), we observe that increase in slip at the interface results in a decrease in the first normal stress difference, $N_1$. The magnitude of $N_1$ is directly dependent on the tensile force acting on the droplet in the direction of shear. While the presence of surfactants, on one hand, enhances the tensile force due to the generation of Marangoni stress, the presence of slip, on the other hand, reduces the magnitude of the Marangoni stress by weakening the interfacial fluid flow. This leads to a decrease in $N_1$ with increase in $\delta$. On the contrary, for $N_2$, increase in $\delta$ is accompanied by an increase in compressive force along the gradient direction due to redistribution of surfactants. This results in a rise in the magnitude of $N_2$, although it has a negative behavior, as seen in fig. 13(b). On comparison between fig. 13(a) and 13(b), it can be observed that the behavior of $N_2$ is just opposite in nature as exhibited by $N_1$. Also, the influence of $k$ on either of $N_1$ and $N_2$ gradually becomes insignificant with an increase in slip, due to reasons specified above. Increase in $\beta$ increases the sensitivity of surface tension towards surfactant distribution and hence a similar nature of variation of the normal stress differences $(N_1, N_2)$ due to change in $\beta$ is expected. For a droplet with higher viscosity, interfacial effects due to alteration of both slip and surfactant concentration become insignificant and hence major changes in $N_1, N_2$ occur in the regime of low viscosity ratio $(\lambda)$ [9,14]. It is worthwhile to mention that, change in magnitude of $N_1$ (due to variation of $\beta$ or $k$) is much larger as compared to a change in $N_2$ for the same variation in $\delta$.

For the other limiting scenario of high $Pe_s$, the effect of slip is observed to be much more pronounced as compared to the limiting case of low $Pe_s$. The total decrease in $N_1$ due to the same increase in $\delta$ is much more enhanced as surface convection is the main mode of surfactant transport. This is evident from the inset in fig. 13(a) which shows the variation in $N_1$ as a function of slip. The variation of $N_2$ with $\delta$, on the contrary, is found to be different altogether (refer to the inset of fig. 13(b)). The magnitude of $N_2$, for this limiting case, is found to decrease with increase in $\delta$. As surfactant transport is more severely affected in this limit due to the presence of slip, both the tensile as well as the compressive components of force acting on the droplet reduces which explains the decrease in the magnitude of first and second normal stress differences $(|N_1|, |N_2|)$.

## VI. CONCLUSION

In this study, we have analyzed the effect of interfacial slip on droplet deformation as well as on the emulsion rheology of a dilute suspension of droplets. Presence of nonlinearity in the system due to consideration of droplet deformation as well as because of the coupled boundary conditions for the flow field and surfactant transport, an exact solution is not possible for arbitrary values of $Pe_s$ or $Ca$. Hence a regular perturbation methodology was adopted with $Ca$ as the perturbation parameter to investigate the small-deformation regime. The present analysis was performed for two different limiting scenarios, namely, high and low $Pe_s$ limit. Although a number of assumptions were made to simplify the problem at hand, never the less, a good match was obtained between our theoretical prediction and the experimental and theoretical results of previous studies. So without exerting any high computational costs, an exact solution was obtained for the limiting cases and a good enough prediction regarding the dynamics of the droplets was possible outside this limiting regime. The results obtained from this analysis put forward some interesting outcomes, which are provided below

(i) Presence of interfacial slip always reduces any deformation of a surfactant-laden droplet suspended in a linear flow (simple shear or a uniaxial extensional flow). The effect of nonuniform surfactant concentration $(k, \beta)$ is seen to be more pronounced for lower values of interfacial slip.

(ii) For the limiting case of high $Pe_s$, a larger decrease in the droplet deformation was observed for the same rise in $\delta$.

(iii) For the same rise in interfacial slip, there is a smaller droplet deformation when the droplet is suspended in a simple shear flow as compared to the case of bulk uniaxial extensional flow.

(iv) For the limiting case of low $Pe_s$, the effective extensional (or shear) viscosity of a droplet suspended in a uniaxial extensional flow (or a simple shear flow) always reduces on account of an increase in the slip at the droplet surface. The role of nonuniform surfactant distribution in increasing the effective viscosity gradually becomes ineffective with an increase in $\delta$. For the other limiting case of $Pe_s \to \infty$, the rate of fall of effective viscosity of the emulsion due to rise in $\delta$ is enhanced.

(v) The normal stress differences $(N_1, N_2)$ are present when the droplet is suspended in a simple shear flow. In the limiting case of low $Pe_s$, $N_1$ is seen to decrease with increase in $\delta$ whereas the magnitude of $N_2$ is found to increase. For the other limiting scenario of high $Pe_s$, the magnitude of both $N_1$ and $N_2$ is seen to reduce with $\delta$.

## Appendix A: Expressions of the constants present in $O(Ca)$ surfactant concentration and shape deformation [in equations (20) and (21)] when the background flow is uniaxial extensional flow

The constants present in the expression of $O(Ca)$ surfactant concentration in equation (20) are provided below

$$\alpha_{2,0}^{(1)} = 16k(1-\beta)(k\beta + 5\lambda - 5\lambda\beta)(16k\beta + 75\lambda - 75\lambda\beta),$$

$$\alpha_{2,0}^{(2)} = k(1-\beta)\begin{pmatrix} 2480\lambda + 512k\beta + 1745\lambda^2 + 548k\lambda\beta - 648k\lambda\beta^2 + 1745\lambda^2\beta^2 \\ +80\beta^2 k^2 + 2480\lambda\beta^2 - 512k\beta^2 + 100k\lambda - 4960\lambda\beta - 3490\lambda^2\beta \end{pmatrix},$$

$$\alpha_{2,0}^{(3)} = k(1-\beta)\begin{pmatrix} 256 + 20k + 4\beta^2 k^2 + 368\lambda - 512\beta + 76\lambda^2 - 152\lambda^2\beta - 5k\lambda\beta \\ +256\beta^2 + 20k\lambda - 736\lambda\beta - 15k\lambda\beta^2 + 76\lambda^2\beta^2 + 40k\beta + 368\lambda\beta^2 - 60k\beta^2 \end{pmatrix},$$

$$\alpha_{2,0}^{(4)} = \{5k\beta + 25\lambda(1-\beta)\}, \quad \alpha_{2,0}^{(5)} = k\beta + 5(1+\lambda)(1-\beta).$$

(A1)

and

$$\alpha_{4,0}^{(1)} = 16k(1-\beta)(32k\beta + 223\lambda - 223\lambda\beta)(k\beta - 5\lambda\beta + 5\lambda),$$

$$\alpha_{4,0}^{(2)} = k(1-\beta)\begin{bmatrix} (-1024k + 6128\lambda + 208k^2 + 6797\lambda^2 - 2088k\lambda)\beta^2 \\ +(1764k\lambda - 12256\lambda - 13594\lambda^2 + 1024k)\beta + 6128\lambda + 6797\lambda^2 + 324k\lambda \end{bmatrix},$$

$$\alpha_{4,0}^{(3)} = k(1-\beta)\begin{bmatrix} (512 + 1120\lambda + 608\lambda^2 - 172k - 187k\lambda + 20k^2)\beta^2 \\ +(-1216\lambda^2 + 151k\lambda - 2240\lambda - 1024 + 136k)\beta \\ +512 + 36k + 1120\lambda + 36k\lambda + 608\lambda^2 \end{bmatrix},$$

(A2)

$$\alpha_{4,0}^{(4)} = \alpha_{2,0}^{(4)}, \quad \alpha_{4,0}^{(5)} = \alpha_{2,0}^{(5)}, \quad \alpha_{4,0}^{(6)} = \{9k\beta + 81\lambda(1-\beta)\}, \quad \alpha_{4,0}^{(7)} = k\beta + 9(1+\lambda)(1-\beta).$$

The constants in the expression for $O(Ca^2)$ shape correction of the droplet [equation (21)] are given by

$$\omega_{2,0}^{(1)} = 4096(k\beta - 5\lambda\beta + 5\lambda)^3,$$

$$\omega_{2,0}^{(2)} = 16(k\beta - 5\lambda\beta + 5\lambda)\begin{Bmatrix} (-2413k\lambda - 768k + 244k^2 + 3840\lambda + 5970\lambda^2)\beta^2 \\ +(768k + 2413k\lambda - 7680\lambda - 11940\lambda^2)\beta + 3840\lambda + 5970\lambda^2 \end{Bmatrix},$$

$$\omega_{2,0}^{(3)} = \begin{bmatrix} \begin{pmatrix} 1104k^3 - 7808k^2 + 12288k - 191280\lambda^2 \\ -61440\lambda - 132630\lambda^3 - 16316k^2\lambda + 80607k\lambda^2 + 77280k\lambda \end{pmatrix}\beta^3 \\ +\begin{pmatrix} 7808k^2 + 16296k^2\lambda + 184320\lambda - 24576k \\ -161214k\lambda^2 + 397890\lambda^3 - 154560k\lambda + 573840\lambda^2 \end{pmatrix}\beta^2 \\ +(12288k + 77280k\lambda - 397890\lambda^3 - 573840\lambda^2 - 184320\lambda + 80607k\lambda^2 + 20k^2\lambda)\beta \\ +61440\lambda + 191280\lambda^2 + 132630\lambda^3 \end{bmatrix},$$

$$\omega_{2,0}^{(4)} = \begin{bmatrix} \begin{pmatrix} 96k^3 - 26583\lambda^2 - 1088k^2 - 11419\lambda^3 \\ -4096 + 3904k + 6964k\lambda^2 + 10832k\lambda - 19152\lambda - 1412k^2\lambda \end{pmatrix}\beta^3 \\ + \begin{pmatrix} 34257\lambda^3 - 13928k\lambda^2 + 1408k^2\lambda \\ +79749\lambda^2 + 1072k^2 - 7808k + 12288 + 57456\lambda - 21664k\lambda \end{pmatrix}\beta^2 \\ + \begin{pmatrix} 16k^2 - 79749\lambda^2 - 12288 - 57456\lambda + 6964k\lambda^2 + 3904k \\ +10832k\lambda + 4k^2\lambda - 34257\lambda^3 \end{pmatrix}\beta \\ +4096 + 26583\lambda^2 + 19152\lambda + 11419\lambda^3 \end{bmatrix}, \quad \text{(A3)}$$

$$\omega_{2,0}^{(5)} = \{5k\beta + 25\lambda(1-\beta)\}, \quad \omega_{2,0}^{(6)} = k\beta + 5(1+\lambda)(1-\beta).$$

and

$$\omega_{4,0}^{(1)} = 10496(k\beta - 9\lambda\beta + 9\lambda)(k\beta - 5\lambda\beta + 5\lambda)^2,$$

$$\omega_{4,0}^{(2)} = 16(k\beta - 5\lambda\beta + 5\lambda)\left\{\begin{matrix}(-1968k - 5444k\lambda + 396k^2 + 16985\lambda^2 + 15088\lambda)\beta^2 \\ +(1968k + 5444k\lambda - 30176\lambda - 33970\lambda^2)\beta + 16985\lambda^2 + 15088\lambda\end{matrix}\right\},$$

$$\omega_{4,0}^{(3)} = \begin{bmatrix} \begin{pmatrix} -22472k^2\lambda + 31488k - 446640\lambda^2 + 1216k^3 - 249586\lambda^3 \\ -199424\lambda - 12672k^2 + 132023k\lambda^2 + 153760k\lambda \end{pmatrix}\beta^3 \\ + \begin{pmatrix} 12672k^2 - 62976k + 748758\lambda^3 - 307520k\lambda + 22436k^2\lambda \\ +598272\lambda + 1339920\lambda^2 - 264046k\lambda^2 \end{pmatrix}\beta^2 \\ + (153760k\lambda + 31488k + 132023k\lambda^2 - 1339920\lambda^2 - 598272\lambda - 748758\lambda^3 + 36k^2\lambda)\beta \\ +199424\lambda + 446640\lambda^2 + 249586\lambda^3 \end{bmatrix}, \quad \text{(A4)}$$

$$\omega_{4,0}^{(4)} = \begin{bmatrix} \begin{pmatrix} -38749\lambda^2 - 1314k^2\lambda - 1208k^2 - 14269\lambda^3 + 72k^3 \\ -10496 + 6336k + 7640k\lambda^2 + 13940k\lambda - 34976\lambda \end{pmatrix}\beta^3 \\ + \begin{pmatrix} -12672k + 42807\lambda^3 + 1200k^2 + 1310k^2\lambda - 27880k\lambda \\ +31488 + 104928\lambda - 15280k\lambda^2 + 116247\lambda^2 \end{pmatrix}\beta^2 \\ + \begin{pmatrix} -104928\lambda + 13940k\lambda - 42807\lambda^3 + 8k^2 - 116247\lambda^2 \\ -31488 + 6336k + 7640k\lambda^2 + 4k^2\lambda \end{pmatrix}\beta \\ +10496 + 34976\lambda + 38749\lambda^2 + 14269\lambda^3 \end{bmatrix},$$

$$\omega_{4,0}^{(5)} = \omega_{2,0}^{(5)}, \quad \omega_{4,0}^{(6)} = \omega_{2,0}^{(6)}, \quad \omega_{4,0}^{(7)} = \{9k\beta + 81\lambda(1-\beta)\}, \quad \omega_{4,0}^{(8)} = k\beta + 9(1+\lambda)(1-\beta)$$

**Appendix B: Expressions of the constant coefficients present in $O(Ca)$ surfactant concentration and shape deformation (equation (29) and (32)) for the case of a simple shear imposed flow**

The constants present in the expression of $O(Ca)$ surfactant distribution on a droplet when suspended in a simple shear flow [equation (29)] are given below

$$\zeta_{2,0}^{(1)} = 16k(1-\beta)(k\beta - 5\lambda\beta + 5\lambda)(16k\beta + 75\lambda - 75\lambda\beta),$$

$$\zeta_{2,0}^{(2)} = k(1-\beta)\begin{bmatrix}(-512k + 80k^2 + 1745\lambda^2 + 2480\lambda - 648k\lambda)\beta^2 \\ +(-4960\lambda + 512k - 3490\lambda^2 + 548k\lambda)\beta \\ +2480\lambda + 1745\lambda^2 + 100k\lambda\end{bmatrix},$$

$$\zeta_{2,0}^{(3)} = k(1-\beta)\begin{bmatrix}(4k^2 + 76\lambda^2 - 15k\lambda + 368\lambda - 60k + 256)\beta^2 \\ +(-5k\lambda - 152\lambda^2 + 40k - 736\lambda - 512)\beta \\ +256 + 20k + 20k\lambda + 368\lambda + 76\lambda^2\end{bmatrix},$$

$$\zeta_{2,0}^{(4)} = 25\lambda(1-\beta) + 5k\beta, \quad \zeta_{2,0}^{(5)} = k\beta + 5(1+\lambda)(1-\beta).$$

(B1)

$$\zeta_{2,2}^{(1)} = k(1-\beta)(400\lambda^2 - 400\beta\lambda^2 + 80\beta k\lambda),$$

$$\zeta_{2,2}^{(2)} = k(1-\beta)\begin{bmatrix}(-400\lambda - 64k\lambda + 64k - 175\lambda^2)\beta \\ +175\lambda^2 + 100k\lambda + 400\lambda\end{bmatrix},$$

$$\zeta_{2,2}^{(3)} = k(1-\beta)\begin{bmatrix}(-4k - 64 - 16k\lambda - 92\lambda - 19\lambda^2)\beta \\ +92\lambda + 20k\lambda + 64 + 19\lambda^2 + 20k\end{bmatrix},$$

$$\zeta_{2,2}^{(4)} = \zeta_{2,0}^{(4)}, \quad \zeta_{2,2}^{(5)} = \zeta_{2,0}^{(5)}.$$

(B2)

$$\zeta_{4,0}^{(1)} = 16k(1-\beta)(32k\beta + 223\lambda - 223\lambda\beta)(k\beta - 5\lambda\beta + 5\lambda),$$

$$\zeta_{4,0}^{(2)} = k(1-\beta)\begin{bmatrix}(-1024k + 208k^2 + 6797\lambda^2 + 6128\lambda - 2088k\lambda)\beta^2 \\ +(-12256\lambda + 1024k - 13594\lambda^2 + 1764k\lambda)\beta \\ +6128\lambda + 6797\lambda^2 + 324k\lambda\end{bmatrix},$$

$$\zeta_{4,0}^{(3)} = k(1-\beta)\begin{bmatrix}(20k^2 + 608\lambda^2 - 187k\lambda + 1120\lambda - 172k + 512)\beta^2 \\ +(151k\lambda - 1216\lambda^2 + 136k - 2240\lambda - 1024)\beta \\ +512 + 36k + 36k\lambda + 1120\lambda + 608\lambda^2\end{bmatrix},$$

$$\zeta_{4,0}^{(4)} = \zeta_{2,2}^{(4)}, \quad \zeta_{4,0}^{(5)} = \zeta_{2,2}^{(5)}, \quad \zeta_{4,0}^{(6)} = 81\lambda(1-\beta) + 9k\beta, \quad \zeta_{4,0}^{(7)} = k\beta + 9(1+\lambda)(1-\beta).$$

(B3)

and

$$\zeta_{4,4}^{(1)} = 16k(1-\beta)(32k\beta + 223\lambda - 223\lambda\beta)(k\beta - 5\lambda\beta + 5\lambda),$$

$$\zeta_{4,4}^{(2)} = k(1-\beta)\begin{bmatrix}(-1024k + 208k^2 + 6797\lambda^2 + 6128\lambda - 2088k\lambda)\beta^2 \\ +(-12256\lambda + 1024k - 13594\lambda^2 + 1764k\lambda)\beta \\ +6128\lambda + 6797\lambda^2 + 324k\lambda\end{bmatrix},$$

$$\zeta_{4,4}^{(3)} = k(1-\beta)\begin{bmatrix}(20k^2 + 608\lambda^2 - 187k\lambda + 1120\lambda - 172k + 512)\beta^2 \\ +(151k\lambda - 1216\lambda^2 + 136k - 2240\lambda - 1024) \\ +512 + 36k + 36k\lambda + 1120\lambda + 608\lambda^2\end{bmatrix},$$

$$\zeta_{4,4}^{(4)} = \zeta_{4,0}^{(4)}, \ \zeta_{4,4}^{(5)} = \zeta_{4,0}^{(5)}, \ \zeta_{4,4}^{(6)} = \zeta_{4,0}^{(6)}, \ \zeta_{4,4}^{(7)} = \zeta_{4,0}^{(7)}.$$

(B4)

The constants present in the expression of $O(Ca^2)$ shape correction in equation (32) are provided below

$$\xi_{2,0}^{(1)} = -102400(k\beta - 5\lambda\beta + 5\lambda)^3,$$

$$\xi_{2,0}^{(2)} = -400(k\beta - 5\lambda\beta + 5\lambda)\left\{\begin{matrix}(-2413k\lambda - 768k + 244k^2 + 3840\lambda + 5970\lambda^2)\beta^2 \\ +(768k + 2413k\lambda - 7680\lambda - 11940\lambda^2)\beta + 3840\lambda + 5970\lambda^2\end{matrix}\right\},$$

$$\xi_{2,0}^{(3)} = \begin{bmatrix}\begin{pmatrix}-1932000k\lambda + 4782000\lambda^2 + 195200k^2 + 3315750\lambda^3 - 307200k \\ +1536000\lambda - 2015175k\lambda^2 + 407900k^2\lambda - 27600k^3\end{pmatrix}\beta^3 \\ +\begin{pmatrix}4030350k\lambda^2 + 3864000k\lambda + 614400k - 14346000\lambda^2 - 195200k^2 \\ -9947250\lambda^3 - 4608000\lambda - 407400k^2\lambda\end{pmatrix}\beta^2 \\ +\begin{pmatrix}-500k^2\lambda + 4608000\lambda - 2015175k\lambda^2 - 1932000k\lambda \\ -307200k + 14346000\lambda^2 + 9947250\lambda^3\end{pmatrix}\beta \\ -4782000\lambda^2 - 3315750\lambda^3 - 1536000\lambda\end{bmatrix},$$

$$\xi_{2,0}^{(4)} = \begin{bmatrix}\begin{pmatrix}35300k^2\lambda + 285475\lambda^3 + 664575\lambda^2 - 174100k\lambda^2 + 102400 \\ +478800\lambda + 27200k^2 - 270800k\lambda - 2400k^3 - 97600k\end{pmatrix}\beta^3 \\ +\begin{pmatrix}-1993725\lambda^2 + 348200k\lambda^2 - 35200k^2\lambda + 195200k \\ +541600k\lambda - 856425\lambda^3 - 1436400\lambda - 26800k^2 - 307200\end{pmatrix}\beta^2 \\ +\begin{pmatrix}1436400\lambda + 1993725\lambda^2 - 100k^2\lambda - 97600k + 856425\lambda^3 \\ -400k^2 - 174100k\lambda^2 - 270800k\lambda + 307200\end{pmatrix}\beta \\ -102400 - 664575\lambda^2 - 285475\lambda^3 - 478800\lambda\end{bmatrix},$$

(B5)

$$\xi_{2,0}^{(5)} = 25\lambda(1-\beta) + 5k\beta, \ \xi_{2,0}^{(6)} = k\beta + 5(1+\lambda)(1-\beta).$$

$$\xi_{2,2}^{(1)} = 80\left\{\left(115k\lambda + 144k - 570\lambda^2 - 720\lambda\right)\beta + 570\lambda^2 + 720\lambda\right\}(k\beta - 5\lambda\beta + 5\lambda),$$

$$\xi_{2,2}^{(2)} = \begin{bmatrix} \begin{pmatrix} -40645k\lambda^2 - 71440k\lambda + 5440k^2 + 4040k^2\lambda - 23040k \\ +99750\lambda^3 + 220800\lambda^2 + 115200\lambda \end{pmatrix}\beta^2 \\ +\left(-441600\lambda^2 - 199500\lambda^3 + 71440k\lambda + 100k^2\lambda + 40645k\lambda^2 + 23040k - 230400\lambda\right)\beta \\ -441600\lambda^2 - 199500\lambda^3 + 71440k\lambda + 100k^2\lambda + 40645k\lambda^2 + 23040k - 230400\lambda \end{bmatrix},$$

$$\xi_{2,2}^{(3)} = \begin{bmatrix} \begin{pmatrix} 34485\lambda^2 - 4465k\lambda^2 - 5440k - 10220k\lambda + 35040\lambda \\ +440k^2\lambda + 11520 + 560k^2 + 10830\lambda^3 \end{pmatrix}\beta^2 \\ +\begin{pmatrix} -21660\lambda^3 + 4465k\lambda^2 - 68970\lambda^2 + 80k^2 + 10220k\lambda \\ -23040 + 20k^2\lambda + 5440k - 70080\lambda \end{pmatrix}\beta \\ +11520 + 34485\lambda^2 + 10830\lambda^3 + 35040\lambda \end{bmatrix},$$

$$\xi_{2,2}^{(4)} = \xi_{2,0}^{(5)}, \ \xi_{2,2}^{(5)} = \xi_{2,0}^{(6)} \tag{B6}$$

$$\xi_{4,0}^{(1)} = 52480(9\lambda + k\beta - 9\lambda\beta)(k\beta - 5\lambda\beta + 5\lambda)^2,$$

$$\xi_{4,0}^{(2)} = 80(k\beta - 5\lambda\beta + 5\lambda)\left\{\begin{matrix}\left(-5444k\lambda - 1968k + 396k^2 + 16985\lambda^2 + 15088\lambda\right)\beta^2 \\ +\left(5444k\lambda + 1968k - 33970\lambda^2 - 30176\lambda\right)\beta + 16985\lambda^2 + 15088\lambda\end{matrix}\right\},$$

$$\xi_{4,0}^{(3)} = \begin{bmatrix} \begin{pmatrix} 157440k - 2233200\lambda^2 - 997120\lambda - 1247930\lambda^3 - 112360k^2\lambda \\ +6080k^3 - 63360k^2 + 768800k\lambda + 660115k\lambda^2 \end{pmatrix}\beta^3 \\ +\begin{pmatrix} 6699600\lambda^2 - 1320230k\lambda^2 - 314880k + 63360k^2 \\ +112180k^2\lambda + 2991360\lambda + 3743790\lambda^3 - 1537600k\lambda \end{pmatrix}\beta^2 \\ +\begin{pmatrix} 660115k\lambda^2 + 180k^2\lambda - 3743790\lambda^3 + 768800k\lambda \\ -2991360\lambda + 157440k - 6699600\lambda^2 \end{pmatrix}\beta \\ +2233200\lambda^2 + 997120\lambda + 1247930\lambda^3 \end{bmatrix}, \tag{B7}$$

$$\xi_{4,0}^{(4)} = \begin{bmatrix} \begin{pmatrix} -6570k^2\lambda - 6040k^2 - 71345\lambda^3 + 360k^3 + 38200k\lambda^2 - 52480 \\ -174880\lambda + 69700k\lambda + 31680k - 193745\lambda^2 \end{pmatrix}\beta^3 \\ + \begin{pmatrix} 6000k^2 + 6550k^2\lambda - 76400k\lambda^2 + 581235\lambda^2 + 524640\lambda \\ +214035\lambda^3 - 63360k + 157440 - 139400k\lambda \end{pmatrix}\beta^2 \\ + \begin{pmatrix} -214035\lambda^3 + 69700k\lambda + 38200k\lambda^2 + 20k^2\lambda - 581235\lambda^2 \\ +40k^2 - 157440 - 524640\lambda + 31680k \end{pmatrix}\beta \\ +52480 + 174880\lambda + 193745\lambda^2 + 71345\lambda^3 \end{bmatrix},$$

$$\xi_{4,0}^{(5)} = \xi_{2,2}^{(4)}, \; \xi_{4,0}^{(6)} = \xi_{2,2}^{(5)}, \; \xi_{4,0}^{(7)} = 81\lambda(1-\beta) + 9k\beta, \; \xi_{4,0}^{(8)} = k\beta + 9(1+\lambda)(1-\beta).$$

and

$$\xi_{4,4}^{(1)} = -52480(9\lambda + k\beta - 9\lambda\beta)(k - 5\lambda\beta + 5\lambda)^2,$$

$$\xi_{4,4}^{(2)} = -80(k\beta - 5\lambda\beta + 5\lambda)\begin{Bmatrix} \left(-5444k\lambda - 1968k + 396k^2 + 16985\lambda^2 + 15088\lambda\right)\beta^2 \\ +\left(5444k\lambda + 1968k - 33970\lambda^2 - 30176\lambda\right)\beta + 16985\lambda^2 + 15088\lambda \end{Bmatrix},$$

$$\xi_{4,4}^{(3)} = \begin{bmatrix} \begin{pmatrix} -157440k + 2233200\lambda^2 + 997120\lambda + 1247930\lambda^3 + 112360k^2\lambda \\ -6080k^3 + 63360k^2 - 768800k\lambda - 660115k\lambda^2 \end{pmatrix}\beta^3 \\ + \begin{pmatrix} -6699600\lambda^2 + 1320230k\lambda^2 + 314880k - 63360k^2 - 112180k^2\lambda \\ -2991360\lambda - 3743790\lambda^3 + 1537600k\lambda \end{pmatrix}\beta \\ + \begin{pmatrix} -660115k\lambda^2 - 180k^2\lambda + 3743790\lambda^3 - 768800k\lambda + 2991360\lambda \\ -157440k + 6699600\lambda^2 \end{pmatrix}\beta \\ -2233200\lambda^2 - 997120\lambda - 1247930\lambda^3 \end{bmatrix}, \quad (B8)$$

$$\xi_{4,4}^{(4)} = \begin{bmatrix} \begin{pmatrix} 6570k^2\lambda + 6040k^2 + 71345\lambda^3 - 360k^3 - 38200k\lambda^2 + 52480 \\ +174880\lambda - 69700k\lambda - 31680k + 193745\lambda^2 \end{pmatrix}\beta^3 \\ + \begin{pmatrix} -6000k^2 - 6550k^2\lambda + 76400k\lambda^2 - 581235\lambda^2 - 524640\lambda \\ -214035\lambda^3 + 63360k - 157440 + 139400k\lambda \end{pmatrix}\beta^2 \\ + \begin{pmatrix} 214035\lambda^3 - 69700k\lambda - 38200k\lambda^2 - 20k^2\lambda + 581235\lambda^2 \\ -40k^2 + 157440 + 524640\lambda - 31680k \end{pmatrix}\beta \\ -52480 - 174880\lambda - 193745\lambda^2 - 71345\lambda^3 \end{bmatrix},$$

$$\xi_{4,4}^{(5)} = \xi_{4,0}^{(5)}, \; \xi_{4,4}^{(6)} = \xi_{4,0}^{(6)}, \; \xi_{4,4}^{(7)} = \xi_{4,0}^{(7)}, \; \xi_{4,4}^{(8)} = \xi_{4,0}^{(8)}.$$

**Appendix C: Expressions of the constants present in equations (38) and (41)**

The constants present in the expression of the effective Trouton viscosity in equation (38) are provided below

$$\varepsilon_1 = 2k\beta + 10\lambda(1-\beta), \varepsilon_2 = k\beta + (5\lambda + 2)(1-\beta),$$

$$\varepsilon_3 = 25\lambda(1-\beta) + 5k\beta, \varepsilon_4 = 5(\lambda+1)(1-\beta) + k\beta,$$

$$\varepsilon_5 = 64(k\beta - 5\lambda\beta + 5\lambda)^3,$$

$$\varepsilon_6 = 4(36k\beta - 48\beta - 179\lambda\beta + 48 + 179\lambda)(k\beta - 5\lambda\beta + 5\lambda)^2,$$

$$\varepsilon_7 = \left[8(k\beta - 5\lambda\beta + 5\lambda)\begin{pmatrix}(-59\lambda k - 36k + 181\lambda + 145\lambda^2 + 24 + 6k^2)\beta^2 \\ +(-48 - 362\lambda + 59\lambda k + 36k - 290\lambda^2)\beta + 181\lambda + 145\lambda^2 + 24\end{pmatrix}\right], \quad (C1)$$

$$\varepsilon_8 = \begin{bmatrix}(-732\lambda + 476\lambda k - 475\lambda^3 - 1179\lambda^2 + 4k^3 - 44k^2 + 144k - 64 + 290k\lambda^2 - 59\lambda k^2)\beta^3 \\ +(-288k + 192 + 3537\lambda^2 - 952\lambda k + 40k^2 + 1425\lambda^3 + 59\lambda k^2 - 580k\lambda^2 + 2196\lambda)\beta^2 \\ +(476\lambda k - 192 - 3537\lambda^2 + 4k^2 - 1425\lambda^3 + 144k - 2196\lambda + 290k\lambda^2)\beta \\ +64 + 1179\lambda^2 + 732\lambda + 475\lambda^3\end{bmatrix}.$$

The constants present in the expression for the normal stress differences, $N_1$ and $N_2$ in equation (41) are provided below

$$\varpi_1^{(1)} = 256(5\lambda + k\beta - 5\lambda\beta)^2,$$

$$\varpi_1^{(2)} = 32(5\lambda + k\beta - 5\lambda\beta)(4k\beta + 16 - 16\beta + 19\lambda - 19\lambda\beta),$$

$$\varpi_1^{(3)} = \begin{bmatrix}(12k^2 + 256 - 152k\lambda + 608\lambda - 128k + 361\lambda^2)\beta^2 \\ +(152k\lambda - 512 - 1216\lambda + 4k^2 - 722\lambda^2 + 128k)\beta \\ +256 + 608\lambda + 361\lambda^2\end{bmatrix}, \quad (C2)$$

$$\varpi_1^{(4)} = 25\lambda(1-\beta) + 5k\beta, \varpi_1^{(5)} = 5(1+\lambda)(1-\beta) + k\beta,$$

and

$$\varpi_2^{(1)} = 8000(5\lambda + k\beta - 5\lambda\beta)^3,$$

$$\varpi_2^{(2)} = 4(4875\lambda + 6000 + 1028k\beta - 4875\lambda\beta - 6000\beta)(5\lambda + k\beta - 5\lambda\beta)^2, \quad (C3)$$

$$\varpi_2^{(3)} = (5\lambda + k\beta - 5\lambda\beta)\begin{bmatrix}(596k^2 - 6976k\lambda - 8224k + 16515\lambda^2 + 38760\lambda + 24000)\beta^2 \\ +(140k^2 + 6976k\lambda + 8224k - 77520\lambda - 33030\lambda^2 - 48000)\beta \\ +16515\lambda^2 + 38760\lambda + 24000\end{bmatrix},$$

$$\varpi_2^{(4)} = \begin{bmatrix} \begin{pmatrix} -5510\lambda^3 - 656k^2 - 19260\lambda - 8000 + 4112k - 599\lambda k^2 \\ +3497k\lambda^2 + 6916k\lambda - 16230\lambda^2 + 24k^3 \end{pmatrix}\beta^3 \\ + \begin{pmatrix} 28k^3 + 57780\lambda - 13832k\lambda - 8224k + 16530\lambda^3 + 48690\lambda^2 \\ +24000 + 576k^2 + 459\lambda k^2 - 6994k\lambda^2 \end{pmatrix}\beta^2 \\ + \begin{pmatrix} -48690\lambda^2 + 6916k\lambda + 4112k + 140\lambda k^2 - \\ 57780\lambda - 16530\lambda^3 - 24000 + 3497k\lambda^2 + 80k^2 \end{pmatrix}\beta \\ +8000 + 19260\lambda + 16230\lambda^2 + 5510\lambda^3 \end{bmatrix},$$

$$\varpi_2^{(5)} = \varpi_1^{(4)}, \quad \varpi_2^{(6)} = \varpi_1^{(5)}.$$

## References

[1] D. Di Carlo, D. Irimia, R.G. Tompkins, and M. Toner, "Continuous inertial focusing, ordering, and separation of particles in microchannels.," Proc. Natl. Acad. Sci. U. S. A., **104**, 18892 (2007).

[2] C.N. Baroud, F. Gallaire, and R. Dangla, "Dynamics of microfluidic droplets.," Lab Chip, **10**, 2032 (2010).

[3] Y. Zhu and Q. Fang, "Analytical detection techniques for droplet microfluidics--a review.," Anal. Chim. Acta, **787**, 24 (2013).

[4] S.-Y. Teh, R. Lin, L.-H. Hung, and A.P. Lee, "Droplet microfluidics.," Lab Chip, **8**, 198 (2008).

[5] X. Li and K. Sarkar, "Effects of inertia on the rheology of a dilute emulsion of drops in shear," J. Rheol. (N. Y. N. Y)., **49**, 1377 (2005).

[6] N. Aggarwal and K. Sarkar, "Rheology of an emulsion of viscoelastic drops in steady shear," J. Nonnewton. Fluid Mech., **150**, 19 (2008).

[7] N. Aggarwal and K. Sarkar, "Deformation and breakup of a viscoelastic drop in a Newtonian matrix under steady shear," J. Fluid Mech., **584**, 1 (2007).

[8] C.L. Tucker III and P. Moldenaers, "Microstructural evolution in polymer blends," Annu. Rev. Fluid Mech., **34**, 177 (2002).

[9] A. Ramachandran and L.G. Leal, "The effect of interfacial slip on the rheology of a dilute emulsion of drops for small capillary numbers," J. Rheol. (N. Y. N. Y)., **56**, 1555 (2012).

[10] R. Pal, "Rheology of simple and multiple emulsions," Curr. Opin. Colloid Interface Sci., **16**, 41 (2011).

[11] G.I. Taylor, "The Viscosity of a Fluid Containing Small Drops of Another Fluid," Proc. R. Soc. A Math. Phys. Eng. Sci., **138**, 41 (1932).

[12] P.M. Vlahovska, J. Bławzdziewicz, and M. Loewenberg, "Small-deformation theory for a surfactant-covered drop in linear flows," J. Fluid Mech., **624**, 293 (2009).

[13] H.A. Stone and L.G. Leal, "The effects of surfactants on drop deformation and breakup," J. Fluid Mech., **220**, 161 (1990).

[14] S. Mandal, S. Das, and S. Chakraborty, "Effect of Marangoni stress on the bulk rheology of a dilute emulsion of surfactant-laden deformable droplets in linear flows," Phys. Rev. Fluids, **2**, 113604 (2017).

[15] W.J. Milliken, H.A. Stone, and L.G. Leal, "The effect of surfactant on the transient motion of Newtonian drops," Phys. Fluids, **5**, 69 (1993).

[16] H.A. Stone, "Dynamics of Drop Deformation and Breakup in Viscous Fluids," Annu. Rev. Fluid Mech., **26**, 65 (1994).

[17] R. Pal, "Fundamental Rheology of Disperse Systems Based on Single-Particle Mechanics," Fluids, **1**, 40 (2016).

[18] R.W. Flumerfelt, "Effects of Dynamic Interfacial Properties on Drop Deformation and Orientation in Shear and Extensional Flow Fields," J. Colloid Interface Sci., **76**, 330 (1980).

[19] S. Velankar, P. Van Puyvelde, J. Mewis, P. Moldenaers, S. Velankar, P. Van Puyvelde, J. Mewis, and P. Moldenaers, "Effect of compatibilization on the breakup of polymeric drops in shear flow," J. Rheol. (N. Y. N. Y)., **45**, 1007 (2001).

[20] Y.T. Hu and A. Lips, "Estimating Surfactant Surface Coverage and Decomposing its Effect on Drop Deformation," Phys. Rev. Lett., **91**, 44501 (2003).

[21] X. Li and C. Pozrikidis, "The effect of surfactants on drop deformation and on the rheology of dilute emulsions in Stokes flow," J. Fluid Mech., **341**, 165 (1997).

[22] S. Yon and C. Pozrikidis, "A Finite-volume/Boundary-element Method for Flow Past Interfaces in the Presence of Surfactants, with Application to Shear Flow Past a Viscous Drop," Comput. Fluids, **27**, 879 (1998).

[23] K. Feigl, D. Megias-Alguacil, P. Fischer, and E.J. Windhab, "Simulation and experiments of droplet deformation and orientation in simple shear flow with surfactants," Chem. Eng. Sci., **62**, 3242 (2007).

[24] R. Zhao and C.W. Macosko, "Slip at polymer–polymer interfaces: Rheological measurements on coextruded multilayers," J. Rheol. (N. Y. N. Y)., **46**, 145 (2002).

[25] P.C. Lee, H.E. Park, D.C. Morse, and C.W. Macosko, "Polymer-polymer interfacial slip in multilayered films," J. Rheol. (N. Y. N. Y)., **53**, 893 (2009).

[26] H.E. Park, P.C. Lee, and C.W. Macosko, "Polymer-polymer interfacial slip by direct visualization and by stress reduction," J. Rheol. (N. Y. N. Y)., **54**, 1207 (2010).

[27] J.L. Goveas and G.H. Fredrickson, "Apparent slip at a polymer-polymer interface," Eur. Phys.

J. B, **92**, 79 (1998).

[28] W. Yu and C. Zhou, "The effect of interfacial viscosity on the droplet dynamics under flow field," J. Polym. Sci. Part B Polym. Phys., **46**, 1505 (2008).

[29] Q. Ehlinger, L. Joly, and O. Pierre-Louis, "Giant Slip at Liquid-Liquid Interfaces Using Hydrophobic Ball Bearings," Phys. Rev. Lett., **110**, 104504 (2013).

[30] A. Ramachandran, K. Tsigklifis, A. Roy, and G. Leal, "The effect of interfacial slip on the dynamics of a drop in flow: Part I. Stretching, relaxation, and breakup," J. Rheol. (N. Y. N. Y)., **56**, 45 (2012).

[31] S. Das, S. Mandal, S.K. Som, and S. Chakraborty, "Effect of interfacial slip on the deformation of a viscoelastic drop in uniaxial extensional flow field," Phys. Fluids, **29**, 32105 (2017).

[32] S. Mandal, A. Bandopadhyay, and S. Chakraborty, "Effect of interfacial slip on the cross-stream migration of a drop in an unbounded Poiseuille flow," Phys. Rev. E, **92**, 23002 (2015).

[33] S. Mandal, S. Das, and S. Chakraborty, "Effect of Marangoni stress on the bulk rheology of a dilute emulsion of surfactant-laden deformable droplets in linear flows," (2017).

[34] L.G. Leal, *Advanced Transport Phenomena* (Cambridge University Press, Cambridge, 2007).

[35] J. Lee and C. Pozrikidis, "Effect of surfactants on the deformation of drops and bubbles in Navier – Stokes flow," Comput. Fluids, **35**, 43 (2006).

[36] S. Das, S. Mandal, and S. Chakraborty, "Cross-stream migration of a surfactant-laden deformable droplet in a Poiseuille flow," Phys. Fluids, **29**, 82004 (2017).

[37] G.K. Batchelor, "The stress system in a suspension of force-free particles," J. Fluid Mech., **41**, 545 (1970).

[38] W.R. Schowalter, C.E. Chaffey, and H. Brenner, "Rheological behavior of a dilute emulsion," J. Colloid Interface Sci., **26**, 152 (1968).